\documentclass{article}
\usepackage{arxiv}

\usepackage[utf8]{inputenc} 
\usepackage[T1]{fontenc}    
\usepackage[hidelinks,breaklinks=true]{hyperref}       
\usepackage{url}            
\usepackage{booktabs}       
\usepackage[version=4]{mhchem}
\usepackage{amsfonts}       
\usepackage{mathtools}
\usepackage{nicefrac}       
\usepackage{lipsum}		
\usepackage{graphicx}
\usepackage[square,numbers]{natbib}
\usepackage{caption}
\usepackage{subcaption}
\usepackage{amsmath}

\usepackage{booktabs}

\usepackage{bm}

\usepackage{url}
\usepackage{siunitx}
\DeclareSIUnit\atmosphere{atm}
\usepackage{pifont}

\usepackage[dvipsnames]{xcolor}
\definecolor{newcolor}{rgb}{.8,.349,.1}
\graphicspath{{Figure/}}

   \newcommand{\mean}[1]{\overline{#1}\,}
   
   \newcommand{\pmean}[1]{\overline{\overline{#1}}\,}
   \newcommand{\logmean}[1]{\overline{#1}^{\text{log}}}

   \newcommand{\hmean}[1]{\overline{#1}^{H}}
   \newcommand{\lmean}[1]{\overline{#1}^{\lambda}}

   \newcommand{\mF}{\mathcal{F}}

\newcommand{\dtp}{\delta^{\,+}}

\newcommand{\dtm}{\delta^{\,-}}

\newcommand{\dd}{\mathrm{d}}

   \usepackage{caption}
   \usepackage{subcaption}
   \usepackage{epsfig,epstopdf}
   \usepackage[normalem]{ulem}
   \newcommand{\beq}{\begin{equation}}
\newcommand{\eeq}{\end{equation}}
\newcommand{\beqs}{\begin{equation*}}
\newcommand{\eeqs}{\end{equation*}}
\usepackage{upgreek}
\title{Pressure-equilibrium-preserving and fully conservative discretization of compressible flow equations for real and thermally perfect gases}%

\date{May 4, 2026}	

\author{
	\href{https://orcid.org/0000-0003-4943-9551}{\includegraphics[scale=0.06]{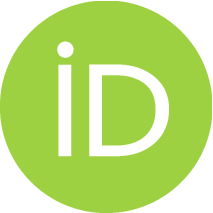}\hspace{1mm}Gennaro Coppola} \\
	Dipartimento di Ingegneria Industriale\\
	Universit\`a di Napoli ``Federico II''\\
	Napoli, Italy \\
	\texttt{gcoppola@unina.it} \\
    \And
\href{https://orcid.org/0009-0003-6376-768X}{\includegraphics[scale=0.06]{orcid.eps}\hspace{1mm} Alessandro {Aiello}}\\
	Dipartimento di Ingegneria Industriale\\
	Universit\`a di Napoli ``Federico II''\\
	Napoli, Italy \\
	\texttt{alessandro.aiello@unina.it} \\
	\And
    \href{https://orcid.org/0000-0002-6518-3114}{\includegraphics[scale=0.06]{orcid.eps}\hspace{1mm} Carlo {De~Michele}}\\
	Gran Sasso Science Institute (GSSI)\\
	L'Aquila, Italy \\
	\texttt{carlo.demichele@gssi.it} \\
}

\renewcommand{\shorttitle}{
PEP and fully conservative discretization for
compressible flows of real and TP gases}%

\hypersetup{
pdftitle={Pressure-equilibrium-preserving and fully conservative discretization of
compressible flow equations for real and thermally perfect gases},
pdfsubject={physics.flu-dyn},
pdfauthor={Gennaro Coppola, Alessandro Aiello, Carlo De~Michele}
pdfkeywords={Compressible flows, Pressure-equilibrium-preserving methods, Kinetic-energy preserving methods, Real gases, Thermally perfect gases},
}
\begin{document}
\maketitle

\begin{abstract}
Numerical simulations of compressible real-fluid flows are notoriously plagued by spurious pressure oscillations arising in regions of abrupt flow variations. As a possible remedy, several numerical formulations enforce the pressure equilibrium condition for the compressible Euler equations, typically at the cost of spoiling the correct conservation of total energy or by overspecifying the thermodynamical variables.
This study proposes for the first time a numerical discretization procedure which is able to discretely preserve the full conservation of the linear invariants (mass, momentum and total energy) and to exactly enforce the pressure equilibrium condition. The method also preserves the conservation of kinetic energy by convection, and is based on the specification of nonlinear numerical fluxes for mass and internal energy which depend on the details of the equation of state. Both thermally perfect and real gases with an arbitrary equation of state are considered, and a simplified approximate pressure equilibrium preserving formulation with excellent performances is also proposed. The effectiveness of the novel formulations is assessed through a series of numerical simulations in supercritical and transcritical conditions with some of the most popular cubic equations of state.
\end{abstract}


\section{Introduction}
The accurate numerical simulation of compressible flows for real and thermally perfect gases is of great importance in a wide array of modern engineering and scientific disciplines.
From the design of high-speed aerospace vehicles and advanced propulsion systems to the analysis of supercritical fluids in power generation and high-pressure turbomachinery, the assumption of a calorically perfect gas frequently breaks down~\cite{Pecnik2012,Jofre2021,Guardone2024}.
In these extreme thermodynamic regimes, the fluid behavior deviates significantly from the ideal gas law, requiring the use of complex, non-linear Equations of State (EoS) to model phenomena such as variable specific heats, dense gas effects, and phase transitions. 

However, the robust discretization of the compressible Euler equations coupled with a generic real gas EoS presents profound numerical challenges. Foremost among these is the generation of spurious, non-physical oscillations across contact discontinuities~\cite{Terashima_JCP_2012}. This phenomenon is fundamentally analogous to the well-documented difficulties encountered in the simulation of multicomponent flows~\cite{Abgrall_JCP_1996}. Just as the abrupt variation of the specific heat ratio across a multi-fluid interface triggers severe numerical artifacts, the highly non-linear dependence of pressure on density and internal energy in a real gas EoS induces similar unphysical fluctuations. If left unmitigated, these spurious oscillations in pressure and velocity can propagate rapidly, leading to non-physical thermodynamic states and ultimately resulting in the catastrophic failure of the simulation.

The root of this issue lies in the extreme difficulty of simultaneously satisfying two fundamental numerical requirements: preserving local pressure equilibrium and maintaining strict, discrete conservation of total energy. 
When fully conservative numerical schemes are applied to real gases, the primary variables (density, momentum and total energy) are updated and exactly conserved. However, the subsequent non-linear inversion of the EoS, required to recover the pressure from the updated density and internal energy, frequently fails to maintain pressure equilibrium across moving contact discontinuities. 
Conversely, many existing strategies attempt to enforce a Pressure-Equilibrium-Preserving (PEP) condition by modifying the energy equation or employing primitive-variable formulations (i.e., discretizing the pressure equation directly in place of the total energy one). While these successfully suppress spurious oscillations, they inherently sacrifice exact discrete conservation of total energy. 
This loss of full conservation can be highly detrimental, as, in the presence of strong shocks, the scheme may fail to converge to the correct weak solution~\cite{Abgrall_JCP_1996,Hou_MC_1994}.

To address the spurious oscillations, a variety of numerical strategies have been proposed over the past decades. Early efforts, predominantly rooted in the study of multicomponent flows, focused on quasi-conservative formulations. As demonstrated by \citet{Karni_JCP_1994}, discretizing the Euler equations in primitive variables avoids the energy-to-pressure EoS inversion, successfully maintaining local pressure equilibrium. To recover partial conservation, \citet{Abgrall_JCP_1996} and later \citet{Shyue_JCP_2001} introduced quasi-conservative schemes wherein the standard conservative fluid equations are solved alongside non-conservative advection equations for the thermodynamic parameters (e.g., the specific heat ratio). By carefully discretizing these auxiliary equations, these methods achieve the PEP property. However, the explicit inclusion of non-conservative terms inherently sacrifices exact discrete conservation of total energy, leading to incorrect shock propagation speeds and thermodynamic anomalies across strong discontinuities.
Similar problems were encountered by evolving the pressure equation for transcritical and supercritical fluids~\cite{Terashima_JCP_2012,Kawai_JCP_2015}, and even the use of enthalpy and internal energy has been attempted~\cite{Lacaze_CF_2019}.

Recognizing the deficiencies of globally non-conservative models, double-flux methods have been introduced to bridge the gap between interface stability and shock fidelity. Originally formalized by \citet{Abgrall_JCP_2001}, this approach was later adapted for real-gas and transcritical flows by \citet{Ma_JCP_2017}. The double-flux strategy employs standard, fully conservative fluxes in the bulk of the flow, but locally switches to a thermodynamically ``frozen", non-conservative auxiliary flux at sharp density gradients to enforce pressure equilibrium. While highly effective at mitigating unphysical waves in practice, double-flux methods still rely on local non-conservation. Furthermore, they require complex and empirical sensor functions to blend the two flux formulations, making them computationally rigid and highly sensitive to tuning parameters when applied to arbitrary, highly non-linear real gases.

To overcome these limitations, efforts were made to modify numerical fluxes directly within a fully conservative framework. For multicomponent flows, \citet{Fujiwara_JCP_2023} sought to address this by deriving equilibrium conditions to construct specialized numerical fluxes without relying on auxiliary transport equations. 
\citet{Terashima_2024} extended this approach to the complex thermodynamic regimes of real gases, developing consistent numerical fluxes maintaining exact conservation of the primary variables. 
However, the final formulation yields only an approximately PEP scheme, with a fixed spatial order of accuracy. As a result, residual pressure oscillations can still manifest when simulating extremely severe thermodynamic gradients or phase transitions.

An alternative philosophy has sought to guarantee exact pressure equilibrium and numerical robustness by coupling PEP conditions with Kinetic-Energy-Preserving (KEP) schemes. The KEP framework was originally developed to prevent non-linear aliasing errors in standard compressible turbulence simulations without relying on numerical dissipation~\cite{Feiereisen_1981, Jameson_JSC_2008, Pirozzoli_JCP_2010}. For calorically perfect ideal gases, researchers have successfully derived discretizations that combine KEP properties with exact pressure equilibrium~\cite{Shima_JCP_2021,Ranocha_CAMC_2021, DeMichele_JCP_2023,DeMichele_JCP_2024,Kawai_JCP_2025}. 
For real-fluid simulations, a recent work by \citet{Bernades_JCP_2023} proposed a specifically designed KEP and PEP scheme. Yet, to strictly maintain both pressure equilibrium and kinetic energy preservation across complex thermodynamic states, they were forced to abandon exact total energy conservation, opting instead to solve a non-conservative pressure evolution equation. Consequently, a numerical methodology that simultaneously guarantees full conservation---particularly of total energy---and exact pressure equilibrium for real and thermally perfect gases remains a missing link in the literature.

To address this critical gap, the present study introduces a novel numerical framework that, for the first time, achieves a fully conservative, kinetic-energy-preserving, and exactly pressure-equilibrium-preserving discretization of the compressible Euler equations for thermally perfect and real gases, with an arbitrary EoS. Our approach demonstrates that it is mathematically and practically possible to retain the exact discrete evolution of the total energy equation without triggering spurious interfacial oscillations, offering a robust and thermodynamically consistent foundation for simulating extreme real-gas phenomena.
To overcome the intrinsic difficulties associated with simulations in transcritical conditions, an approximate formulation is also proposed, based on a slight modification of the set of exact fluxes. By retaining excellent PEP properties, with unrivaled performances among existing formulations, the set of approximate fluxes shows robust behavior also in extreme conditions.

The remainder of this paper is organized as follows. Section~\ref{sec:problem_formulation} establishes the problem formulation, detailing the continuous compressible Euler equations and the underlying discretization framework. In Section~\ref{sec:PEP_Discretization}, we present the core methodology, rigorously deriving the fully conservative numerical fluxes and proving their KEP and PEP properties at the discrete level. This generic formulation is subsequently particularized for specific thermodynamic models: Section~\ref{sec:thermally_perfect} addresses thermally perfect gases, while Section~\ref{sec:real_gas} extends the approach to real gases, with a specific focus on the van der Waals and Peng--Robinson equations of state. Section~\ref{sec:results} presents a suite of numerical experiments---including a standard density wave test and a multidimensional 
mixing case---designed to validate both the PEP property and exact discrete conservation. Finally, Section~\ref{sec:conclusions} summarizes the key findings and offers concluding remarks.

\section{Problem formulation}\label{sec:problem_formulation}
\subsection{Governing equations}
The compressible Euler equations, which express the conservation of mass, momentum and total energy for compressible inviscid flows, can be written as:
\begin{align}
\dfrac{\partial \rho}{\partial t} &= -\dfrac{\partial \rho u_{\alpha}}{\partial x_{\alpha}} \;, \label{eq:Mass} \\[3pt]
\dfrac{\partial \rho u_{\beta}}{\partial t} &= -\dfrac{\partial \rho u_{\alpha}u_{\beta}}{\partial x_{\alpha}} -\dfrac{\partial p}{\partial x_{\beta}}  \;, \label{eq:Momentum} \\[3pt]
\dfrac{\partial \rho E}{\partial t} &=  -\dfrac{\partial \rho u_{\alpha}E}{\partial x_{\alpha}} - \dfrac{\partial pu_{\alpha}}{\partial x_{\alpha}},
\; \label{eq:TotEnergy}
\end{align}
where $\rho$ is the density, $u_{\alpha}$ is the Cartesian velocity component along $x_\alpha$, $p$ is the pressure and $E$ the
total energy per unit mass, which is the sum of the internal ($e$) and kinetic ($\kappa = u_{\alpha}u_{\alpha}/2$) energies per unit mass: $E = e + \kappa$.
We will assume the convention that Greek subscripts, such as $\alpha$ or $\beta$, refer to the components of Cartesian vectors, whereas Latin subscripts, such as $i$ or $j$ appearing in the subsequent sections of this paper, are used to denote the values of the discretized variable on a nodal point. Unless otherwise stated, the summation convention over repeated Greek indices is assumed.
The system \eqref{eq:Mass}--\eqref{eq:TotEnergy} is closed by an equation of state, relating $p,\rho$ and the temperature $T$, and by specifying the dependence of internal energy on temperature and density, which is typically provided through suitable departure functions.

In what follows, we will also need to consider the induced balance equations
for additional quantities related to the primary variables $\rho$, $\rho u_{\beta}$, and $\rho E$.
These equations are easily derived
by combining Eqs.~\eqref{eq:Mass}--\eqref{eq:TotEnergy}
and by applying the chain and product rules of differentiation,
with respect to both temporal and spatial variables.
In particular, we will use the evolution equations for the kinetic and internal energies:
\begin{align}
\dfrac{\partial \rho \kappa}{\partial t} &= -\dfrac{\partial \rho u_{\alpha} \kappa}{\partial x_{\alpha}} - u_\alpha\dfrac{\partial p}{\partial x_{\alpha}},\label{eq:KinEnergy}\\
\dfrac{\partial \rho e}{\partial t} &= -\dfrac{\partial \rho u_{\alpha} e}{\partial x_{\alpha}} -p\dfrac{\partial u_{\alpha}}{\partial x_{\alpha}},\label{eq:IntEnergy}
\end{align}
whose sum returns the total energy equation \eqref{eq:TotEnergy}.
Moreover, to discuss the PEP property, which is the main subject of the present paper, we will use the evolution equations for 
the velocity components $u_\beta$, and the pressure $p$, which can be written as
\begin{align}
\dfrac{\partial u_{\beta}}{\partial t} &= -u_{\alpha}\dfrac{\partial u_{\beta}}{\partial x_{\alpha}} - \dfrac{1}{\rho}\dfrac{\partial p}{\partial x_{\beta}},\label{eq:Velocity} \\
\dfrac{\partial p}{\partial t} &=  -\dfrac{\partial p u_{\alpha}}{\partial x_{\alpha}} -\left(\rho c^2 -p\right)\dfrac{\partial u_{\alpha}}{\partial x_{\alpha}},\label{eq:Pressure}
\end{align}
where $c = \sqrt{(\partial p/\partial\rho)_s}$ is the speed of sound, $s$ being the specific entropy.
Eqs.~\eqref{eq:Velocity} and \eqref{eq:Pressure} show that if at a given time instant a spatially constant state is assumed for both $u_\beta$ and $p$, all the spatial derivatives at the right-hand sides vanish. This also causes the temporal derivatives at the left-hand side to vanish, meaning that the constant distribution of pressure and velocity remains constant also in time. This is the Pressure Equilibrium property for the compressible Euler equations, which allows for the existence of density wave solutions traveling at constant velocity in a uniform pressure field.

\subsection{Discretization setting}
Our focus is on the discretization of the spatial terms in Eqs.~\eqref{eq:Mass}--\eqref{eq:TotEnergy}, for which we always use central (non-dissipative) schemes. To isolate spatial errors, we assume in the theoretical analysis that the manipulation of time derivatives can be performed as in the continuous case. This type of semidiscretized analysis leads to a system of Ordinary Differential Equations (ODE) whose temporal integration is performed by using standard solvers.
In this paper, the theory will be developed for the one-dimensional version of Eq.~\eqref{eq:Mass}--\eqref{eq:TotEnergy}, which is obtained by omitting the Greek subscripts. The extension to the general three-dimensional case is straightforward and involves simply including the analogous spatial terms along the additional Cartesian components.
The analysis is developed by referring to a Finite Difference (FD) treatment of the convective terms in Eqs.~\eqref{eq:Mass}--\eqref{eq:TotEnergy} on a uniform mesh $\left\{x_i\right\}$ of width $h=x_{i+1}-x_i$, although most of our results could also be applied in other frameworks, such as Finite Volume or Finite Element methods, and can be generalized to nonuniform or curvilinear meshes, following standard approaches~\cite{Pirozzoli_JCP_2011,Kuya_JCP_2021}. Finally, we focus on the discretization of the various spatial operators at internal points, neglecting the effects of boundary conditions, which could be studied with ad hoc methods.

Although we work in a FD framework, we will design the discretization of the convective spatial terms in Eqs.~\eqref{eq:Mass}--\eqref{eq:TotEnergy} by specifying the numerical fluxes, i.e. by assuming a \textit{locally conservative} discretization in the form
\begin{equation}\label{eq:Diff_of_Fluxes}
    \left.\dfrac{\partial\rho u\phi}{\partial x}\right|_i \approx \dfrac{1}{h}\left(\mF_{\rho\phi}^{i+\frac{1}{2}}-\mF_{\rho\phi}^{i-\frac{1}{2}}\right)
\end{equation}
where $\phi$ is a generic transported quantity per unit mass ($1$, $u$ or $E$ in Eq.~\eqref{eq:Mass}, \eqref{eq:Momentum} or \eqref{eq:TotEnergy}, respectively).
The numerical flux 
$\mF_{\rho\phi}^{i+\frac{1}{2}}$ is a consistent approximation of $\rho u\phi$ at $x_{i+1/2}=x_i+h/2$, and is a function of the sets of nodal values $\rho_i, u_i$ and $\phi_i$. 
Note that, even if the right-hand sides of Eq.~\eqref{eq:Mass}--\eqref{eq:TotEnergy} can be expressed as the divergence of a single term including convective and pressure mechanisms, Eq.~\eqref{eq:Diff_of_Fluxes} refers to the specification of the numerical fluxes for the convective contribution only (the first terms at the r.h.s.~of Eqs.~\eqref{eq:Mass}--\eqref{eq:TotEnergy}).
The terms involving the pressure in Eq.~\eqref{eq:Momentum} and \eqref{eq:TotEnergy} are discretized by using standard central derivative formulas, as specified in Section~\ref{sec:KEP_Discretization}.
Eq.~\eqref{eq:Diff_of_Fluxes} reproduces the divergence structure of the convective terms in Eqs.~\eqref{eq:Mass}--\eqref{eq:TotEnergy} and implies also global conservation of the invariant $\rho\phi$ by virtue of the telescoping property.

To simplify the notation, we will make use of the difference and average operators
\begin{equation*}
    \dtm\varphi_i = \varphi_i-\varphi_{i-1},\qquad\dtp\varphi_i = \varphi_{i+1}-\varphi_i,\qquad \mean{\varphi}_i = \left(\varphi_{i+1}+\varphi_i\right)/2
\end{equation*}
which allows to write the central difference as $\varphi_{i+1}-\varphi_{i-1} = 2\dtm\mean{\varphi}_i$ and to express Eq.~\eqref{eq:Diff_of_Fluxes} as
\begin{equation*}\label{eq:Diff_of_Fluxes_short}
    \left.\dfrac{\partial\rho u\phi}{\partial x}\right|_i \approx \dfrac{1}{h}\dtm\mF_{\rho\phi}^{i+\frac{1}{2}}.
\end{equation*}
In most cases, we will omit the apex when referring to the flux evaluated at $x_{i+1/2}$: $\mF_{\rho\phi}=\mF_{\rho\phi}^{i+\frac{1}{2}}$, and, when no ambiguity can arise, we also omit the suffix in the expressions of the difference or of the average of a generic quantity: $\mean{\phi} = \mean{\phi}_i, \dtm\phi=\dtm\phi_i$.

Within the described setting, the spatial discretization of the r.h.s.~of Eqs.~\eqref{eq:Mass}--\eqref{eq:TotEnergy} is completely determined by the specification of the convective fluxes $\mF_\rho, \mF_{\rho u},\mF_{\rho E}$ and by the discretization details of the pressure terms $\partial p/\partial x$ and $\partial up/\partial x$ in Eqs.~\eqref{eq:Momentum} and \eqref{eq:TotEnergy}, respectively.
The derivation exposed in the next sections considers second-order approximations, which implies that the numerical flux $\mF_{\rho\phi}$ is in general a two-point interpolation of $\left(\rho u\phi\right)_{i+1/2}$, and is a function of $\rho,u$ and $\phi$ at nodes $x_i$ and $x_{i+1}$ only.
The extension to higher orders is obtained by using standard arguments and is detailed in several references~\cite{LeFloch_SIAMJNA_2002,Fisher_JCP_2013b,Ranocha_JSC_2018,DeMichele_JCP_2024,Aiello_JCP_2025}.

\subsection{Kinetic Energy Preserving formulation}\label{sec:KEP_Discretization}
A first constraint on the possible choice for the numerical fluxes is given by the requirement that the formulation satisfies the Kinetic Energy Preserving (KEP) property~\cite{Feiereisen_1981,Jameson_JSC_2008,Coppola_AMR_2019,Coppola_JCP_2019}, which amounts to the requirement that the discrete induced evolution equation for the kinetic energy has a convective term that is locally and globally conservative, as for the continuous equation (cf.~Eq.~\eqref{eq:KinEnergy}). It is well known (\cite{Jameson_JSC_2008,Veldman_JCP_2019,Coppola_JCP_2023}) that, for second order fluxes, the KEP property is satisfied when the convective fluxes for mass and momentum are linked by the simple relation
\begin{equation*}\label{eq:KEP_condition}
    \mF_{\rho u} = \mF_\rho\,\mean{u}.
\end{equation*}
In this case, kinetic energy evolves according to a discrete equation analogous to Eq.~\eqref{eq:KinEnergy}, in which the convective term can be cast as a difference of numerical fluxes, with flux~\cite{Coppola_JCP_2023,DeMichele_C&F_2023}
\begin{equation}\label{eq:KinEnFlux}
   \mF_{\rho\kappa}=\mF_{\rho}\dfrac{u_iu_{i+1}}{2} .
\end{equation}
The requirement that the discrete total energy evolves coherently with the induced evolution of kinetic energy suggests designing the convective numerical flux for total energy $\mF_{\rho E}$ as the sum of the kinetic-energy flux given by Eq.~\eqref{eq:KinEnFlux} and of a convective flux for internal energy $\mF_{\rho e}$.
The conservative pressure term is correspondingly built as the sum of the induced discretized pressure term in Eq.~\eqref{eq:KinEnergy} (which belongs to the discrete pressure term in the momentum equation \eqref{eq:Momentum}) and of a consistent discretization of the pressure term in Eq.~\eqref{eq:IntEnergy}. 
Assuming standard second-order central schemes for these terms, one has 
\begin{equation}\label{eq:PressureTerms_Discrete}
    \left.u_i\dfrac{\partial p}{\partial x}\right|_i \approx u_i\dfrac{1}{h}\dtm\mean{p}\qquad\qquad \left.p_i\dfrac{\partial u}{\partial x}\right|_i\approx p_i\dfrac{1}{h}\dtm\mean{u}
\end{equation}
which gives the correct discretization of the pressure term in the total energy equation as consistent with the advective form of the derivative of the product:
\begin{equation*}\label{eq:PressureTerms_Discrete_2}
    \left.\dfrac{\partial up}{\partial x}\right|_i =u_i\left.\dfrac{\partial p}{\partial x}\right|_i + p_i\left.\dfrac{\partial u}{\partial x}\right|_i\approx \dfrac{1}{h}\left(u_i\dtm\mean{p}+p_i\dtm\mean{u}\right) = \dfrac{1}{h}\dtm\pmean{up}
\end{equation*}
where $\pmean{up} = \left(u_ip_{i+1}+p_iu_{i+1}\right)/2$ is the product mean~\cite{Coppola_JCP_2019}.

In conclusion, a locally conservative and KEP discretization is characterized by the set of total fluxes (i.e. including both convective and pressure contributions) given by
\begin{equation}\label{eq:KEP_Flux_Tot}
\mathcal{F}_{\rho}^{\text{tot}}= \mathcal{F}_{\rho},\qquad\qquad
\mathcal{F}_{\rho u}^{\text{tot}} =\mathcal{F}_{\rho}\,\mean{u}+\mean{p},\qquad\qquad
\mathcal{F}_{\rho E}^{\text{tot}} = \mathcal{F}_{\rho e}+\mathcal{F}_{\rho}\dfrac{u_iu_{i+1}}{2}+\pmean{up},
\end{equation}
where the mass and internal energy fluxes $\mF_\rho$ and $\mF_{\rho e}$ are still unspecified, except for the obvious requirement that they are consistent approximations of mass and internal energy fluxes. 
The set of fluxes reported in Eq.~\eqref{eq:KEP_Flux_Tot} defines a formulation which is locally conservative of linear invariants $\rho, \rho u$ and $\rho E$ and preserves the (local and global) conservation of kinetic energy by convection. It embodies many KEP formulations available in the literature that have gained popularity in recent years.
As examples, the simple choice $\mF_\rho = \mean{\rho}\,\mean{u}$ and $\mF_{\rho e} = \mean{\rho}\,\mean{u}\,\mean{e}$ gives the KEEP scheme~\cite{Kuya_JCP_2018} which, even if it does not have exact additional structural properties, has been widely used because of its good performances for ideal gases. The choice $\mF_\rho = \mean{\rho}\,\mean{u}$ and $\mF_{\rho e} = \mean{\rho}\,\mean{u}\,\hmean{e}$ (where  $\hmean{e} = 2e_ie_{i+1}/(e_i+e_{i+1})$ is the harmonic mean) gives the zeroth-order AEC scheme, developed in~\cite{DeMichele_JCP_2023}, whereas 
$\mF_\rho = \mean{\rho}\,\mean{u}$ and $\mF_{\rho e} = \mean{\rho e}\,\mean{u}$ give the KEEP$_\text{PE}$ scheme~\cite{Shima_JCP_2021}, these last two formulations being PEP for ideal gases.
Recently, the present authors also determined mass and internal energy fluxes such that the formulation in Eq.~\eqref{eq:KEP_Flux_Tot}  is Entropy Conservative (EC), i.e. it is able to discretely preserve the correct balance of entropy, for the case of real gases with an arbitrary equation of state~\cite{Aiello_JCP_2025} and for thermally perfect gases~\cite{Aiello_ArXiv_2025}. In the next section, we derive an expression for $\mF_\rho$ and $\mF_{\rho e}$ which gives a formulation able to satisfy the PEP condition for thermally perfect and real gases with an arbitrary EoS.

\section{PEP formulation}\label{sec:PEP_Discretization}
In this section, we analyze the general problem of enforcing the PEP condition for the system of compressible Euler equations, without making any assumptions about the equation of state. In Sections~\ref{sec:thermally_perfect} and \ref{sec:real_gas} we explicitly work out the particular cases of a thermally perfect gas and of various thermodynamic models for real gases. 

\subsection{Theoretical derivation of exact PEP fluxes}\label{sec:Theoretical_PEP}
In order to discretely enforce the pressure equilibrium property, it is necessary to consider the discrete evolution equations for velocity components and pressure, which are the induced discrete counterparts of Eqs.~\eqref{eq:Velocity} and \eqref{eq:Pressure}. 
Among these, the pressure evolution equation is the more challenging one, due to its explicit dependence on the EoS. We therefore begin with the discrete velocity equation, which is independent of the EoS. General conditions for the discrete enforcement of the velocity equilibrium have previously been derived for the case of calorically perfect gases~\cite{DeMichele_JCP_2024,Ranocha_CAMC_2021}, where it was shown that the KEP condition is sufficient to enforce it. This result still applies but, for completeness, we briefly revisit the derivation here.

First of all, we explicitly write the semidiscrete evolution equations for mass and momentum as obtained by discretizing Eqs.~\eqref{eq:Mass} and \eqref{eq:Momentum} according to the prescriptions illustrated in the previous section:
\begin{align}
\dfrac{\dd \rho_i}{\dd t} &= -\dfrac{1}{h}\dtm\mF_\rho\;, \label{eq:Mass_Discr} \\[3pt]
\dfrac{\dd \rho_i u_i}{\dd t} &= -\dfrac{1}{h}\dtm\mF_\rho\mean{u} - \dfrac{1}{h}\dtm\mean{p} \;. \label{eq:Momentum_Discr} 
\end{align}
The discrete evolution equation for the velocity can be now obtained by using the expression of the time derivative of the velocity $u$ in terms of the time derivatives of $\rho$ and $\rho u$
\begin{equation}
    \frac{\partial u}{\partial t} =
    \frac{1}{\rho}\frac{\partial \rho u}{\partial t} -
    \frac{u}{\rho} \frac{\partial \rho}{\partial t}.
\end{equation}
From this relation, substituting Eqs.~\eqref{eq:Mass_Discr} and \eqref{eq:Momentum_Discr}, one obtains
\begin{equation}
    \dfrac{\dd u_i}{\dd t} =
    -\dfrac{1}{\rho_ih}\left(\dtm\mF_\rho\mean{u} + \dtm\mean{p} -
    u_i\,\dtm\mF_\rho\right).
\end{equation}
In the case of spatially constant pressure $p_i = P$ and velocity $u_i = U$, the pressure contribution at the right-hand side vanishes, and one is left with
\begin{equation}\label{eq:velocity_equilibrium_condition}
    \dfrac{\dd u_i}{\dd t} =
    -\dfrac{1}{\rho_ih}\left(U\dtm\hat\mF_\rho -
    U\dtm\hat\mF_\rho\right)=0,
\end{equation}
where we denoted with $\hat{\mF}_{\rho}$ the form assumed by $\mF_\rho$ for uniform velocity and pressure fields.
Eq.~\eqref{eq:velocity_equilibrium_condition} shows that for the KEP discretization adopted, a uniform spatial distribution of pressure and velocity always induces a zero time derivative for $u_i$, which implies that the velocity field remains uniform as time evolves.

Moving on to the discrete evolution of pressure, we start by considering the internal energy per unit volume $\rho e$ and use an equation of state in the form $\rho e = \rho e(\rho,p)$.
By taking the time derivative of this relation, we get
\beqs
\dfrac{\partial\rho e}{\partial t} = \left(\dfrac{\partial\rho e}{\partial\rho}\right)_p\dfrac{\partial\rho}{\partial t}+\left(\dfrac{\partial\rho e}{\partial p}\right)_\rho\dfrac{\partial p}{\partial t}.
\eeqs
Substituting the temporal derivatives with the right-hand sides of the mass and internal energy equations \eqref{eq:Mass} and \eqref{eq:IntEnergy}, we obtain
\beqs
\dfrac{\partial \rho ue}{\partial x}+ p\dfrac{\partial u}{\partial x}= \alpha(\rho,p)\dfrac{\partial\rho u}{\partial x}+\beta(\rho,p)\dfrac{\partial p}{\partial t},
\eeqs
where we defined 
\beqs
\alpha(\rho,p) = \left(\dfrac{\partial\rho e}{\partial\rho}\right)_p \qquad\text{and}\qquad
\beta(\rho,p) = \left(\dfrac{\partial\rho e}{\partial p}\right)_\rho.
\eeqs 
Now we assume to discretize the convective terms in the mass and internal energy equations conservatively, i.e. as the difference of numerical fluxes $\dtm\mF/h$, and to use the simple central discretization for the pressure term in the internal energy equation given in Eq.~\eqref{eq:PressureTerms_Discrete}.
This gives
\beqs
\dfrac{1}{h}\delta^-\mF_{\rho e} + p_i\,\dfrac{1}{h}\dtm\mean{u}= \alpha(\rho_i,p_i)\dfrac{1}{h}\delta^-\mF_\rho+\beta(\rho_i,p_i)\dfrac{\dd p_i}{\dd t}.
\eeqs
Assuming again uniform spatial distributions of pressure and velocity, one is left with
\beq\label{eq:Pcond_inst}
\delta^-\hat{\mF}_{\rho e}= \alpha(\rho_i,P)\,\delta^-\hat{\mF}_\rho+h\,\beta(\rho_i,P)\dfrac{\dd p_i}{\dd t},
\eeq
where, as usual, $\hat{\mF}_{\rho e}$ and $\hat{\mF}_\rho$ denote the form assumed by the numerical fluxes for uniform velocity and pressure fields.
Eq.~\eqref{eq:Pcond_inst} reveals that to discretely maintain the uniform pressure distribution (i.e. to enforce the condition $\dd p_i/\dd t = 0$), the mass and internal energy fluxes have to satisfy the constraint
\beq\label{eq:Pcond}
\delta^-\hat{\mF}_{\rho e}= \hat{\alpha}_i\,\delta^-\hat{\mF}_\rho,
\eeq
where $\hat{\alpha}_i = \hat{\alpha}(\rho_i) = \alpha(\rho_i,P)$.

A condition analogous to that in Eq.~\eqref{eq:Pcond} has already been analyzed in \cite{Fujiwara_JCP_2023} for the case of a multicomponent mixture of real gases and has been the starting point to obtain approximate PEP formulations in \cite{Terashima_2024}.
An exact enforcement of Eq.~\eqref{eq:Pcond} seems problematic at first sight, since it requires the specification of the numerical fluxes $\mF_\rho$ and $\mF_{\rho e}$ such that the product between the exact difference $\delta^-\hat\mF_\rho$ and an arbitrary function $\hat\alpha(\rho)$ can be expressed as the difference of numerical fluxes $\delta^-\hat\mF_{\rho e}$. 
In Eq.~\eqref{eq:Pcond}, the function $\hat\alpha(\rho)$ is fixed by the thermodynamic model, whereas $\hat\mF_\rho$ and $\hat\mF_{\rho e}$ can be chosen arbitrarily, provided that they are consistent approximations of the mass and internal energy fluxes.

To solve the problem expressed by Eq.~\eqref{eq:Pcond} we proceed inspired by the steps of the theory of Tadmor~\cite{Tadmor_MC_1987,Tadmor_AN_2003} and observe that $\hat\alpha_i\,\delta^-\hat\mF_\rho$ is an exact difference if (and only if) $\hat\mF_\rho\,\delta^+\hat\alpha_i$ is, since the sum $a_i\delta^-b_i+b_i\delta^+a_i$ (which is a consistent discretization of the advective form of the derivative of the product $ab$) admits the decomposition as a difference of fluxes
$a_i\delta^-b_i+b_i\delta^+a_i = \delta^-(a_{i+1}b_i).$
This allows us to rewrite Eq.~\eqref{eq:Pcond} in the equivalent form
\beqs
\dtm\hat\mF_{\rho e} = - \hat\mF_\rho\dtp\hat\alpha_i+\dtm\left(\hat\alpha_{i+1}\hat\mF_\rho\right),
\eeqs
and we are now faced with the problem of finding a numerical flux $\mF_\rho$ such that $\hat\mF_\rho\dtp\hat{\alpha}_i$ is an exact difference.
We proceed now by observing that the flux $\mF_\rho$ should be a consistent approximation of the analytical flux $f_\rho(\rho,u) = \rho u$ at $x_{i+1/2}$.
We assume that the function $\hat{\alpha}(\rho)$ is (at least locally) invertible, in such a way that the inverse function $\rho(\hat{\alpha})$ can be considered, and the relation $\rho\,$--$\,\hat{\alpha}$ is a local one-to-one mapping. This confers to the variable $\hat{\alpha}$ the role of an {\it entropy} variable, in the terminology of the theory of Tadmor.
In this case we can express the function $\hat{f}_\rho(\rho) = f_\rho(\rho,U)$ through the new variable $\hat{\alpha}$ by defining the function $\hat{g}_\rho(\hat{\alpha}) = \hat{f}_\rho(\rho(\hat{\alpha}))$,
which is approximated by the numerical flux $\hat{\mathcal{G}}_{\rho}(\hat{\alpha}_i,\hat{\alpha}_{i+1})= \hat\mF_\rho(\rho(\hat{\alpha}_i),\rho(\hat{\alpha}_{i+1}))$.

To express now $\hat{\mathcal{G}}_{\rho}\dtp\hat{\alpha}$ as an exact  difference, we 
make use of the primitive function $\uppsi(\hat{\alpha})$ defined by
\beqs \label{eq:DefPsi}
\uppsi(\hat{\alpha}) = \int\hat{g}_\rho(\hat{\alpha})\,\dd\hat{\alpha}.
\eeqs 
Using the Integral Mean Value Theorem, we can write
\beq \label{eq:DeltaPsi}
\delta^+\uppsi_i = \int_{\hat{\alpha}_i}^{\hat{\alpha}_{i+1}}\hat{g}_\rho(\hat{\alpha})\,\dd\hat{\alpha} = \hat{g}_\rho(\hat{\alpha}^*)\,\delta^+\hat{\alpha}
\eeq 
where $\hat{\alpha}^*$ is a unknown value between $\hat{\alpha}_i$ and $\hat{\alpha}_{i+1}$ and the value $\hat{g}_\rho(\hat{\alpha}^*)$ is a second-order approximation of $\hat{g}_{\rho}(\hat{\alpha}_{i+\frac{1}{2}})$.
Eq.~\eqref{eq:DeltaPsi} suggests the adoption of the value
$\hat{\mathcal{G}}_\rho=\hat{g}_{\rho}(\hat{\alpha}^*)$, i.e.
\beqs \label{eq:Fflux_alpha}
\hat{\mathcal{G}}_\rho(\hat{\alpha}_i,\hat{\alpha}_{i+1}) = \dfrac{\delta^+\uppsi_i}{\delta^+\hat{\alpha}_i},
\eeqs 
which, expressed in the variables $\rho_i$ gives:
\beq \label{eq:Frho_flux}
\hat{\mathcal{F}}_\rho(\rho_i,\rho_{i+1}) =\hat{\mathcal{G}}_\rho(\hat{\alpha}(\rho_i),\hat{\alpha}(\rho_{i+1})) = \dfrac{\delta^+\psi_i}{\delta^+\hat{\alpha}_i}
\eeq 
where ${\psi}(\rho) = \uppsi(\hat{\alpha}(\rho))$.
This gives:
\begin{equation}\label{eq:PCond_hat}
\hat{\alpha}_i\delta^-\hat{\mF}_\rho=-\hat{\mF}_\rho\delta^+\hat{\alpha}_i + \delta^-\left(\hat{\alpha}_{i+1}\hat{\mF}_\rho\right) = -\delta^+\psi_i+\delta^-\left(\hat{\alpha}_{i+1}\hat{\mF}_\rho\right) = \delta^-\left(\hat{\alpha}_{i+1}\hat{\mF}_\rho-\psi_{i+1}\right),
\end{equation}
where we used $\dtp\psi_i = \dtm\psi_{i+1}$.
Eq.~\eqref{eq:PCond_hat} reduces to Eq.~\eqref{eq:Pcond} with the choice $\hat{\mF}_{\rho e} = \hat{\alpha}_{i+1}\hat{\mF}_\rho-\psi_{i+1}$.
The form of $\hat{\mF}_{\rho e}$ can be made more symmetric by observing that $\hat{\alpha}_{i+1} = \mean{\hat{\alpha}}_i + \delta^+\hat{\alpha}_i/2$
which implies 
\beqs
\hat{\alpha}_{i+1}\hat{\mF}_\rho = \mean{\hat{\alpha}}_i\hat{\mF}_\rho + \dfrac{\hat{\mF}_\rho\delta^+\hat{\alpha}_i}{2} = \mean{\hat{\alpha}}_i\hat{\mF}_\rho + \dfrac{\delta^+\psi_i}{2}
\eeqs
 eventually giving
\beq\label{eq:Frhoe_flux}
\hat{\mF}_{\rho e} = \mean{\hat{\alpha}}_{i}\hat{\mF}_\rho-\mean{\psi}_{i}.
\eeq 
Equations \eqref{eq:Frho_flux} and \eqref{eq:Frhoe_flux} satisfy Eq.~\eqref{eq:Pcond}, and constitute the basis to build the general fluxes $\mF_\rho$ and $\mF_{\rho e}$ enforcing the PEP property.

To show how the function $\psi$ and the fluxes in Eqs.~\eqref{eq:Frho_flux} and \eqref{eq:Frhoe_flux} can be practically calculated, let us define the function
\beq
\lambda(\rho,p) = \left(\dfrac{\partial e}{\partial \rho}\right)_p
\eeq
which is related to $\alpha$ by the equation $\alpha = e+\rho\lambda$.
By using $f_\rho(\rho,u) = \rho u$ the function $\psi(\rho)$ can be calculated as
\beq\label{eq:psi_of_rho}
\psi(\rho) = \uppsi(\alpha(\rho)) = \int\hat{g}_\rho(\hat\alpha(\rho))\,\dd\hat\alpha(\rho) =
\int \hat{f}_\rho (\rho)\hat\alpha'\,\dd\rho = 
U\int\rho\left(2\hat\lambda+\rho\hat\lambda'\right)\dd\rho = U\rho^2\hat\lambda
\eeq
where primes denote differentiation and, as usual, $\hat\lambda(\rho) = \lambda(\rho,U)$. 
Eq.~\eqref{eq:psi_of_rho} allows us to write the fluxes satisfying Eq.~\eqref{eq:Pcond} in the explicit form
\begin{equation}\label{eq:PEP_Fluxes_hat}
    \hat\mF_\rho =  \dfrac{\delta^+\rho_i^2\hat\lambda_i}{\delta^+\hat\alpha_i}U\qquad\qquad
    \hat\mF_{\rho e} = \mean{\hat\alpha}\,\hat\mF_\rho-U\,\mean{\rho^2\hat\lambda} 
\end{equation}
The final form of $\mF_\rho$ and $\mF_{\rho e}$ can now be obtained in several ways, all reducing to Eq~\eqref{eq:PEP_Fluxes_hat} for uniform velocity.
In what follows we use the simple extension of Eq.~\eqref{eq:PEP_Fluxes_hat} obtained by adopting the arithmetic mean for $u$, leading to:
\begin{equation}\label{eq:PEP_Fluxes}
    \mF_\rho = \lmean{\rho}\mean{u}\qquad\qquad
    \mF_{\rho e} = \mean{\alpha}\,\mF_\rho-\mean{u}\,\mean{\rho^2\lambda}. 
\end{equation}
where we defined the average $\mean{\rho}^\lambda=\delta^+(\rho_i^2\lambda_i)/\delta^+\alpha_i$.

\subsection{Treatment of the singularity and approximate PEP formulation}

Equations~\eqref{eq:PEP_Fluxes} and \eqref{eq:KEP_Flux_Tot} furnish the final formulation satisfying the KEP and PEP properties for an arbitrary EoS, whose details are embodied in the functions $\alpha(\rho,p)$ and $\lambda(\rho,p)$. 
As it is typical in these cases, the nonlinear average $\lmean{\rho}$ in the mass flux is potentially singular for uniform distributions of $\alpha$.
A similar situation occurs, for example, in the EC and PEP (for ideal gases) flux by Ranocha~\cite{Ranocha_2020}, which employs the logarithmic mean ($\logmean{\phi} = \dtp\phi/\dtp\log\phi)$, and in the EC flux for real gases recently developed in~\cite{Aiello_JCP_2025}, which also uses a potentially singular flux for uniform distributions of temperature. 
This phenomenon, which is fundamentally linked to the definition of the average through the use of the integral mean value theorem~\cite{DeMichele_JCP_2025}, can cause severe limitations in the applications, especially when large regions of uniform distributions of temperature or density are expected to occur.
To avoid the singularity, suitable fixes can be devised, typically implemented by locally reverting to non-singular numerical fluxes when the denominators of the singular averages ($\dtp\alpha_i$ in our case) fall under a specific tolerance.
These non-singular schemes can be either a suitable Taylor expansion of the original mean, when possible, as in the case of the logarithmic mean used in entropy conservative methods for ideal gases (\cite{Ismail_JCP_2009,Ranocha_2020,DeMichele_JCP_2023,Kawai_JCP_2025}) or standard non-singular schemes chosen among those with good performances, to minimize errors, as in \cite{Aiello_JCP_2025}.
Recently, a more advanced technique based on the theory of discrete gradient operators has been also used~\cite{Klein2026}.

In the present case, since a full series expansion of the singular mean $\lmean{\rho}$ appearing in the mass flux $\mF_\rho$ seems cumbersome for an arbitrary EoS, we choose to use the simple modification of the flux $\mF_\rho$ in Eq.~\eqref{eq:PEP_Fluxes} obtained by adopting the arithmetic mean $\mean{\rho}$ in place of $\lmean{\rho}$, i.e. by using the fluxes 
\begin{equation}\label{eq:APEP_Fluxes}
    \mF_\rho = \mean{\rho}\,\mean{u}\qquad\qquad
    \mF_{\rho e} = \mean{\alpha}\,\mF_\rho-\mean{u}\,\mean{\rho^2\lambda}. 
\end{equation}
The set of fluxes given by Eqs.~\eqref{eq:KEP_Flux_Tot} and \eqref{eq:APEP_Fluxes} actually shows excellent performances on highly challenging tests, as reported in Section~\ref{sec:results}.
These results tempted us to use it not only as a fix for the exact PEP formulation given by Eq.~\eqref{eq:PEP_Fluxes}, but also as an approximate PEP formulation which can be used during the whole simulation, irrespective of the local value of $\dtp\alpha_i$.
Similar very good performances have also been observed by using other classical algebraic means for the density in the mass flux (e.g.~the harmonic or geometric means), suggesting that the form of the interpolation for the density in the mass flux has only marginal effects on the enforcement of the PEP property. 
For the sake of simplicity, we adopt here the arithmetic mean, leaving a thorough analysis of this subject as a possible future work.
In conclusion, although only approximately PEP, the excellent performances of the formulation \eqref{eq:KEP_Flux_Tot}--\eqref{eq:APEP_Fluxes} induce us to offer it as a sufficiently simple formulation which could be used in place of Eqs.~\eqref{eq:KEP_Flux_Tot}--\eqref{eq:PEP_Fluxes}.
We will refer to the Exact PEP formulation (EPEP-RG) for the scheme given by Eqs.~\eqref{eq:KEP_Flux_Tot}--\eqref{eq:PEP_Fluxes}, and to the Approximate PEP formulation (APEP-RG) for that given by Eqs.~\eqref{eq:KEP_Flux_Tot}--\eqref{eq:APEP_Fluxes}.

As a final comment, we observe that in the case of a calorically perfect gas, for which $\rho e = p/(\gamma -1)$, the function $\alpha$ is identically zero, and the whole derivation exposed in Section~\ref{sec:Theoretical_PEP}, that led to Eq.~\eqref{eq:PEP_Fluxes}, breaks down. However, even if the mass flux in Eq.~\eqref{eq:PEP_Fluxes} remains undefined, the internal energy flux can be safely determined, and turns out to reduce to $\mF_{\rho e} =\mean{\rho e}\,\mean{u}$, which is the internal energy flux of the formulation $\text{KEEP}_\text{PE}$ by \citet{Shima_JCP_2021}. 
Hence, the APEP-RG formulation defined by the fluxes in Eq.~\eqref{eq:APEP_Fluxes} nicely reduces to the  $\text{KEEP}_\text{PE}$ formulation in the case of a calorically perfect gas.

\section{Thermally perfect gases}\label{sec:thermally_perfect}
In this section, we work out the particular case of a thermally perfect gas model, for which the usual perfect-gas EoS is assumed: $p = \rho R T$, where $R$ is the gas constant and $T$ is the absolute temperature. The internal energy depends on temperature  through temperature-variable isochoric specific heat capacity $c_v(T)$, which implies 
\begin{equation}\label{eq:IntEnergy_TP}
e =\int_{T_{\mathrm{ref}}}^{T} c_v(T')\, \dd{T'} + e_{\text{ref}}
\end{equation}
and the ‘‘ref" subscript indicates some reference condition. 
We use a polynomial-based approach for the modeling of $c_v(T)$,~\cite{Hansen_SANDIA_2019}, by which the isochoric specific heat is expressed using temperature-based polynomial fittings:
\begin{equation}\label{eq:cv_poly}
    {c}_v(T) = \sum_{k=0}^N c_k {T}^k.
\end{equation}
This functional dependence is widely used when thermal equilibrium is assumed~\cite{Hansen_SANDIA_2019}, and is also easily tractable from an analytical point of view. 
We will detail the derivation for the general formulation with arbitrary $N$, although only 5, 7 or 9 coefficients are typically used to experimentally fit the gas behaviour~\cite{NASA_CHASE,NASA}. 
By substituting Eq.~\eqref{eq:cv_poly} into Eq.~\eqref{eq:IntEnergy_TP} and using the perfect-gas EoS one  has
\begin{equation}\label{eq:alpha_lambda_TP}
\alpha(\rho,p) = -\sum_{k = 1}^N A_kT^{k+1}+ \varepsilon_{\text{ref}},\qquad
\lambda(\rho,p) = -\sum_{k = 1}^N c_k\dfrac{T^{k+1}}{\rho} 
\end{equation}
with $A_k = kc_k/(k+1)$ and $T(\rho,p) = p/\rho R$. The final form of the PEP fluxes corresponding with the fluxes in Eq.~\eqref{eq:PEP_Fluxes} for a thermally perfect gas is 
\begin{equation}\label{eq:PEP_Fluxes_TP}
    \mF_\rho = \dfrac{\dtp\sum_{k=1}^Nc_k\rho T^{k+1}}{\dtp\sum_{k=1}^NA_kT^{k+1}}\mean{u}\qquad\qquad
    \mF_{\rho e} = -\mean{\left(\sum_{k=1}^NA_kT^{k+1}\right)}\,\mF_\rho+\mean{u}\,\mean{\left(\sum_{k=1}^Nc_k\rho T^{k+1}\right)}. 
\end{equation}

As for the general form of the fluxes in Eq.~\eqref{eq:PEP_Fluxes}, the mass flux for thermally perfect gases in Eq.~\eqref{eq:PEP_Fluxes_TP} is potentially singular. However, the particular polynomial form assumed for the $c_v(T)$ allows one to derive a formulation that is singularity-free in all conditions, without the need for a fix.
To obtain this formulation, we need to derive the general form of the mass flux by starting from the PEP condition in Eq.~\eqref{eq:PEP_Fluxes_hat}, which, using Eq.~\eqref{eq:alpha_lambda_TP}, can be written, for thermally perfect gases, as
\begin{equation*}
    \hat{\mF}_\rho = \dfrac{\sum_{k=1}^Nc_k\dtp T^{k}}{\sum_{k=1}^NA_k\dtp T^{k+1}} \dfrac{PU}{R}
\end{equation*}
where we used the ideal gas equation of state particularized to the case of uniform pressure: $\rho = P/RT$.
Using now the general identity
\begin{equation*}
    \dtp \phi^k = \dtp \phi \sum_{j=0}^{k-1} \phi^{k-1-j}_{i} \phi^j_{i+1},
\end{equation*}
one finally gets
\begin{equation*}
    \hat{\mF}_\rho
    = \dfrac{\sum_{k=1}^Nc_k \sum_{j=0}^{k-1} T^{k-1-j}_{i} T^j_{i+1}}{\sum_{k=1}^NA_k\sum_{j=0}^{k} T^{k-j}_{i} T^j_{i+1}} \dfrac{PU}{R},
\end{equation*}
for which the denominator does not vanish. The final form of the flux can now be obtained by assuming an arithmetic interpolation for pressure and velocity:
\begin{equation}\label{eq:PEP_MassFlux_TP2}
    \mF_\rho
    = \dfrac{\sum_{k=1}^Nc_k S_i^k}{\sum_{k=1}^NA_kS_i^{k+1}} \dfrac{\mean{p}\,\mean{u}}{R}
\end{equation}
where $S_i^k=\sum_{j=0}^{k-1} T^{k-1-j}_{i} T^j_{i+1}$. The flux can be calculated more efficiently by observing the identity $S_i^{k+1} = T_iS_i^k + T_{i+1}^k$.
The mass flux in Eq.~\eqref{eq:PEP_MassFlux_TP2}, together with the internal energy flux in Eq.~\eqref{eq:PEP_Fluxes_TP}, is a more efficient PEP formulation for thermally perfect gases, and is the one adopted in the numerical tests in Section~\ref{sec:results}.
Note that in practical applications, the polynomial fitting for the $c_v$ in Eq.~\eqref{eq:cv_poly} can also include terms with negative exponents. The treatment can be easily generalized to this case as in~\cite{Aiello_ArXiv_2025}, and the fluxes in Eqs.~\eqref{eq:PEP_Fluxes_TP} and \eqref{eq:PEP_MassFlux_TP2} are consequently adapted.

\section{Real gases}\label{sec:real_gas}
For real gases, we adopt the usual representation of the internal energy by means of departure functions, whose expressions are given, for example, in \cite{Tosun} assuming the form
\begin{equation}\label{eq:IntEnergy_RG}
e(\rho,T) =e^{\text{TP}}(T) + D^e(\rho,T)=  \int_{T_{\mathrm{ref}}}^{T} c_v\left(T'\right)\, \dd{T'} + e_{\text{ref}} +D^e(\rho,T), 
\end{equation}
where $e^{\text{TP}}$ is the thermally perfect contribution given in Eq.~\eqref{eq:IntEnergy_TP} and $D^e$ is the suitable departure function for internal energy~\cite{Tosun}. In general, we are denoting with $D^\phi$ the departure function for the thermodynamic quantity $\phi$, which is defined by $D^\phi=\phi-\phi^{\text{TP}}$ and $\phi^{\text{TP}}$ is the thermally perfect contribution. 

Standard manipulations give 
\begin{equation}
\lambda(\rho,p)=\left(\frac{\partial e}{\partial \rho}\right)_p=\left(\frac{\partial e^\text{TP}}{\partial \rho}\right)_p+\left(\frac{\partial D^e}{\partial \rho}\right)_p=c_v(T)\left(\frac{\partial T}{\partial \rho}\right)_p + \left(\frac{\partial D^e}{\partial\rho}\right)_T+\left(\frac{\partial D^e}{\partial T}\right)_\rho \left(\frac{\partial T}{\partial \rho}\right)_p,
\end{equation}
where $T = T(\rho,p)$ is the explicit form of the equation of state.
Gathering common terms and noting that, by definition, $(\partial D^e/\partial T)_\rho=D^{c_v}(\rho,T)$, yields
\begin{equation}\label{eq:alpha_lambda_RG}
\lambda(\rho,p) = c_v^{\text{RG}}(\rho,T)\left(\dfrac{\partial T}{\partial \rho}\right)_p + \left(\dfrac{\partial D^e}{\partial \rho}\right)_T,
\qquad
\alpha(\rho,p) = e+\rho\left(c_v^{\text{RG}}(\rho,T)\left(\dfrac{\partial T}{\partial \rho}\right)_p + \left(\dfrac{\partial D^e}{\partial \rho}\right)_T\right).
\end{equation}
 
Finally, for the evaluation of  the speed of sound $c$, required by the use of the pressure evolution equation in the main Euler system, we will always use the standard expression (particularized for each equation of state)
\begin{equation}
    c^2 = \frac{1}{\beta_s} = \frac{1}{\beta_T}+\frac{T\left[\left(\partial p/\partial T\right)_\rho\right]^2}{\rho c_v^{\text{RG}}},\qquad \text{with}\qquad\beta_T=\frac{1}{\rho\left(\partial p/\partial \rho\right)_T}
\end{equation}
$\beta_s$, $\beta_T$ being the isentropic and isothermal compressibility, respectively.

In the following sections, specializations for the van der Waals and Peng--Robinson models of Eq.~\eqref{eq:alpha_lambda_RG} will be derived in non-dimensional form. 

\subsection{Van der Waals model}
The van der Waals model is taken into account due to its simplicity as 
a first simple real-gas correction to the ideal gas equation of state.
The equation of state for pressure reads 
\begin{equation}
    p =\frac{\rho T}{1-\rho b} - a\rho^2
\end{equation}
with $a^*=(27/64)(R^*T^*_c)^2/p^*_c$ and $b^*=(1/8)R^*T^*_c/p^*_c$, with the superscript $^*$ indicating dimensional values and the suffix `$c$' referring to critical conditions.
Internal energy is given by
\begin{equation}
    e = e^{\text{TP}} + D^e = \int_{T_{\mathrm{ref}}}^T c_v(T')\,\mathrm{d}T' -a\rho
\end{equation}
with $D^e=-a\rho$. With such definitions, Eq.~\eqref{eq:alpha_lambda_RG} becomes
\begin{equation}
\lambda(\rho,p)=\left(c_v(T)\left(a-\frac{p}{\rho^2} -2a\rho b\right) - a\right),\qquad\alpha(\rho,p)=e + \rho\left(c_v(T)\left(a-\frac{p}{\rho^2} -2a\rho b \right) - a\right),
\end{equation}
where $c_v(T)=c_v^{\text{RG}}(T)$, since $D^{c_v}=\int_0^\rho\rho^{-2}T(\partial^2_{TT}p)|_\rho\,\mathrm{d}\rho=0$.

\subsection{Peng--Robinson model}
The Peng--Robinson model is considered because of its widespread application in high-pressure/low-temperature flows, as it overcomes the intrinsic instability of the van der Waals model near the critical region. The equation of state for pressure is
\begin{equation}
    p =\frac{\rho T}{(1-\rho b)} - \frac{\rho^2aA(T)}{1 + 2\rho b - (\rho b)^2}
\end{equation}
with
\beq
    A(T)=\left(1 + k\left(1-\sqrt{\frac{TT^*_{\mathrm{ref}}}{T_c^*}}\right)\right)^2,\qquad
    k=0.37464 + 1.54226\omega - 0.26992\omega^2 
\eeq
and  $a^*=0.45724(R^*T^*_c)^2/p^*_c$, $b^*=0.0778(R^*T^*_c/p^*_c)$. $A(T)$ is the Soave function for accounting temperature dependence on potential energy, while $\omega=0.2249$ is the acentric factor for CO$_2$.
Internal energy has the expression
\begin{equation}
    e = e^{\text{TP}} + D^e = \int_{T_{\mathrm{ref}}}^T c_v(T')\,\mathrm{d}T' -\frac{a(TA'(T)-A(T))}{2\sqrt{2}b}\log{\left(\frac{1+(1+\sqrt{2})\rho b}{1+(1-\sqrt{2})\rho b}\right)}.    
\end{equation}
For this specific model, $D^{c_v}\ne 0$, and is
\begin{equation}
D^{c_v}=\int_0^\rho\rho^{-2}T\left(  \frac{\partial^2p}{\partial T^2}\right)_\rho\,\mathrm{d}\rho=\frac{TaA''(T)}{2\sqrt{2}b}\log{\left(\frac{1+(1+\sqrt{2})\rho b}{1+(1-\sqrt{2})\rho b}\right)}.
\end{equation}
The expression for $\lambda(\rho,p)$ is calculated by using Eq.~\eqref{eq:alpha_lambda_RG} with
\begin{equation*}
 \left(\frac{\partial T}{\partial \rho}\right)_p=-\dfrac{\left(\partial p/\partial \rho\right)_T}{\left(\partial p/\partial T\right)_\rho}\quad\text{and}\quad
    \left(\frac{\partial D^e}{\partial \rho}\right)_T=\frac{a(TA'(T)-A(T))}{1 + 2\rho b - (\rho b)^2}.
\end{equation*}
\section{Numerical results}\label{sec:results}
\begin{table}
    \centering
    {
    \begin{tabular}{c c c c c c c}
        \textbf{Scheme} &  \textbf{Ref.} & $\boldsymbol{\mathcal{F}_\rho}$ & $\boldsymbol{\mathcal{F}_{\rho e}}$& 
        \textbf{PEP IG} &  \textbf{PEP RG} & \textbf{Fully Cons.}         
        \vspace{0.1cm}\\
        \hline \\
        EPEP-RG & new  & $\lmean{\rho}\mean{u}$ & $\mean{\alpha}\mF_\rho-\mean{u}\mean{\rho^2\lambda}$ & n.a. &\ding{51} & \ding{51}\\
    \vspace{0.cm}\\
        APEP-RG & new  & $\mean{\rho}\,\mean{u}$ & $\mean{\alpha}\mF_\rho-\mean{u}\mean{\rho^2\lambda}$ & \ding{51} &\ding{109} & \ding{51}\\
        \vspace{0.cm}\\
        APEC & \citet{Terashima_2024}  & $\mean{\rho}\,\mean{u}$ & $\left(\mean{\rho e}-\frac{\dtp\alpha\,\dtp\rho}{4}\right)\mean{u}$ & \ding{51} &\ding{109} & \ding{51}\\     
    \vspace{0.cm}\\
        KEEP & \citet{Kuya_JCP_2018}  & $\mean{\rho}\,\mean{u}$ & $\mF_\rho\,\mean{e}$ & \ding{55} &\ding{55} & \ding{51}\\         \vspace{0.cm}\\
        $\text{KEEP}_\text{PE}$ & \citet{Shima_JCP_2021}  & $\mean{\rho}\,\mean{u}$ & $\mean{\rho e}\,\mean{u}$ & \ding{51} &\ding{55} & \ding{51}\\        
    \vspace{0.cm}\\
        $\text{KGP}_\text{Pt}$ & \citet{Bernades_JCP_2023}  & $\mean{\rho}\,\mean{u}$ & -- & \ding{51} &\ding{51} & \ding{55}\\          
    \vspace{0.cm}\\
        \hline\\        
    \end{tabular}}
    \caption{Summary of the compared numerical discretizations. \ding{51}: property verified, {\ding{109}}: property verified \textit{approximately}, \ding{55}: property not verified, n.a. : not applicable. $\lambda = (\partial e/\partial\rho)_p$, $\alpha = (\partial \rho e/\partial\rho)_p$, $\lmean{\rho} =\dtp(\rho^2\lambda)/\dtp\alpha$. Momentum and total energy fluxes are calculated according to Eq.~\eqref{eq:KEP_Flux_Tot}. 
 }
    \label{tab:schemes}
\end{table}

In this section, we present numerical tests designed to assess the proposed discrete formulations. The focus in on verifying the correct fulfillment of the PEP property and the exact total-energy conservation in inviscid flows.
To place the performance of the proposed schemes in context, we compare them against classical formulations from the literature; a summary of all considered schemes in real gases simulations is provided in Table~\ref{tab:schemes}.
For the thermally perfect gas simulations, our formulation employs the mass flux in Eq.~\eqref{eq:PEP_MassFlux_TP2} and the internal energy flux contained in Eq.~\eqref{eq:PEP_Fluxes_TP}.
To compare the performances of our newly derived schemes, we consider the KEEP scheme by~\citet{Kuya_JCP_2018}, commonly used for simulations of calorically perfect gases, and its variant KEEP$_{\text{PE}}$ by~\citet{Shima_JCP_2021} which is PEP for calorically perfect gases; regarding formulations specifically designed for real-gas simulations we consider the Approximately Pressure Equilibrium Conserving (APEC) scheme developed by~\citet{Terashima_2024} and the pressure-based KGP$_{\text{Pt}}$ scheme studied in~\citet{Bernades_JCP_2023}.
In all tests, time integration is carried out using the standard fourth-order Runge--Kutta method.

We consider the compressible Euler equations in dimensionless form: reference quantities are set as the standard ambient conditions for temperature and pressure (SATP), i.e.~$T^*_{\mathrm{SATP}}=\SI{298.15}{\kelvin}$ and $p^*_{\mathrm{SATP}}=\SI{1}{\atmosphere}$.
 This normalization enables a consistent treatment of different thermodynamic regimes and allows the equation of state to be adapted to conditions representative of each model. In particular, we examine thermally perfect gases at high enthalpy, van der Waals fluids at supercritical conditions, and Peng--Robinson fluids at both supercritical and transcritical regimes.

The assessment is carried out using two benchmark problems. A one-dimensional density wave is used to evaluate the PEP property and to compare the proposed schemes against classical counterparts. Additionally, a two-dimensional double-jet configuration is employed to investigate the schemes’ behavior in a more demanding flow setting, studying both the energy conservation and the insurgence of non-physical pressure oscillations.

\subsection{One-dimensional density wave}
\begin{figure}[tb]
    \centering
    \begin{subfigure}[b]{0.33\textwidth}
    \centering
    \includegraphics[width=\textwidth]{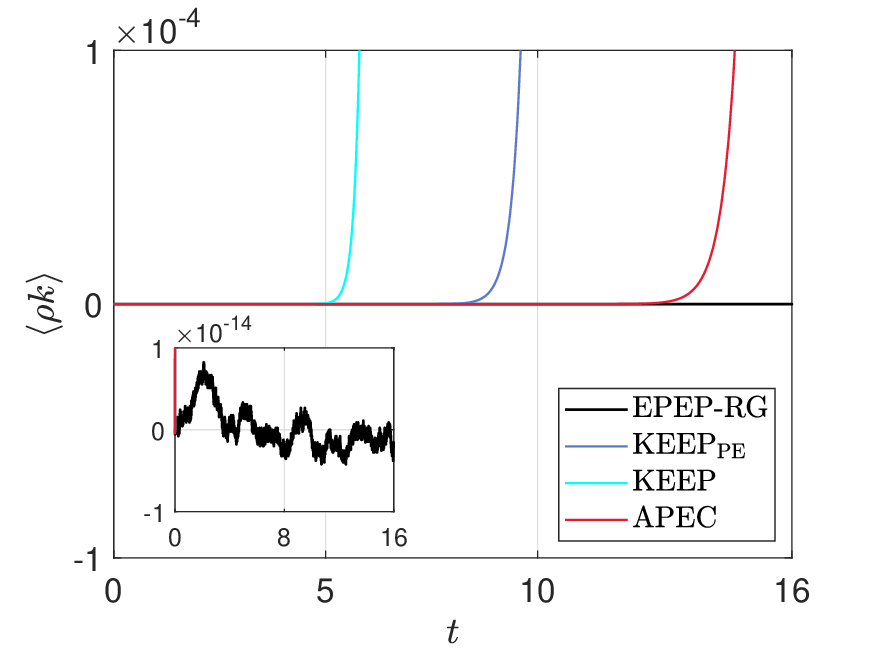}
    \caption{Thermally perfect gas\\\phantom{a}}
    \label{fig:DW_rhok_TP}
\end{subfigure}
    \begin{subfigure}[b]{0.33\textwidth}
    \centering
    \includegraphics[width=\textwidth]{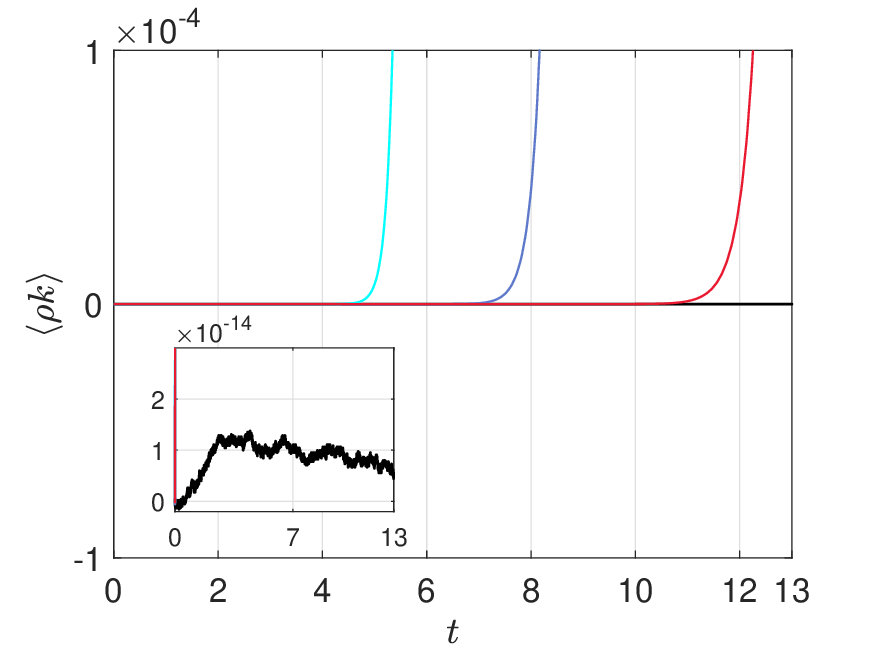}
    \caption{Van der Waals EoS (supercritical \\conditions)}
    \label{fig:DW_rhok_VDW}
\end{subfigure}
    \begin{subfigure}[b]{0.33\textwidth}
    \centering
    \includegraphics[width=\textwidth]{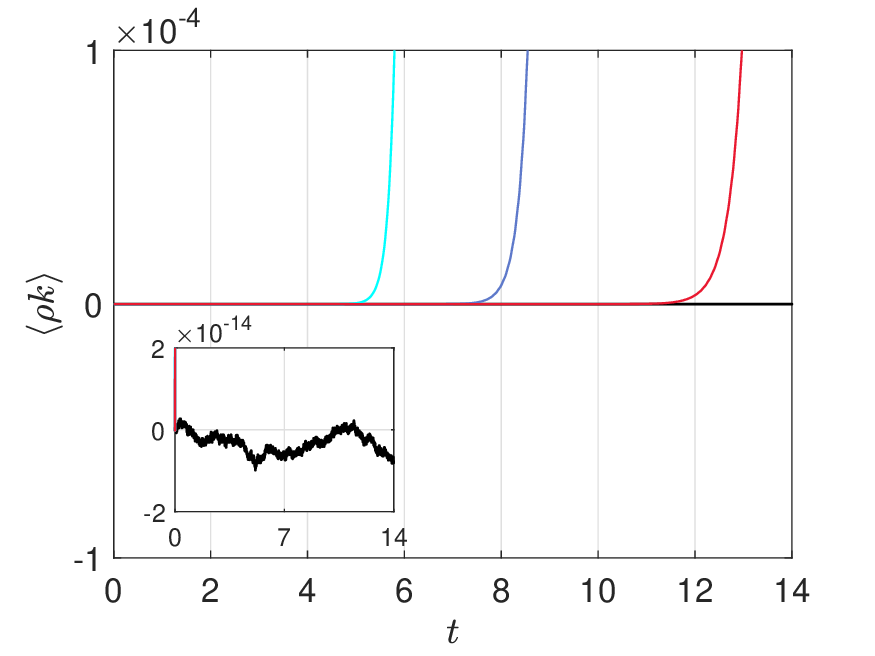}
    \caption{Peng--Robinson EoS (supercritical \\conditions).}
    \label{fig:DW_rhok_PR}
\end{subfigure}
    \caption{Global kinetic-energy evolution for the one-dimensional density wave test with various gas models.}\label{fig:DW_rhok}    
\end{figure}
\begin{figure}[p]
    \centering
    \begin{subfigure}[b]{0.45\textwidth}
    \centering
    \includegraphics[width=\linewidth]{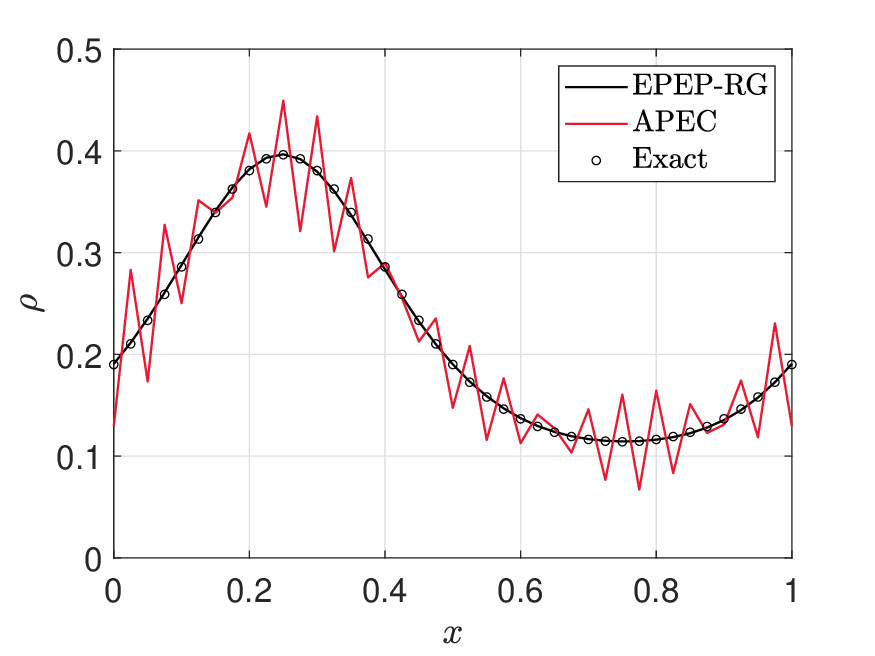}
    \caption{}
    \end{subfigure}
    \begin{subfigure}[b]{0.45\textwidth}
    \centering
    \includegraphics[width=\linewidth]{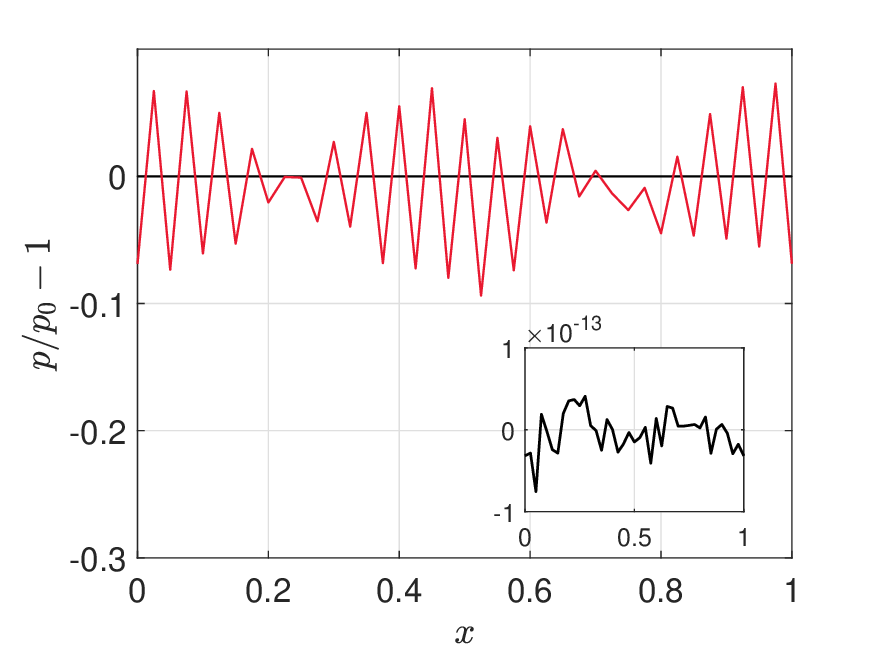}
    \caption{}
    \end{subfigure}
    \begin{subfigure}[b]{0.45\textwidth}
    \centering
    \includegraphics[width=\linewidth]{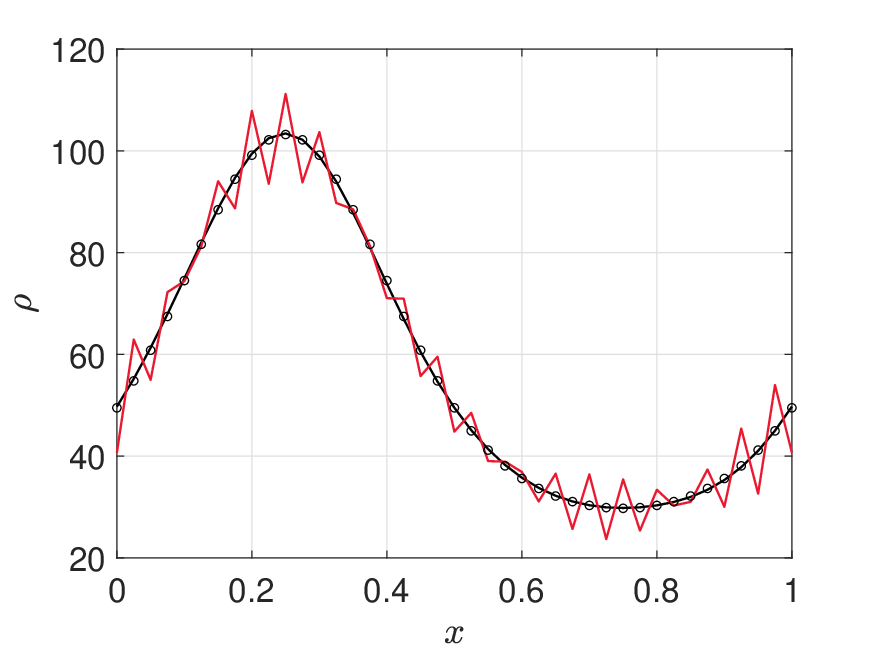}
    \caption{}
    \end{subfigure}
    \begin{subfigure}[b]{0.45\textwidth}
    \centering
    \includegraphics[width=\linewidth]{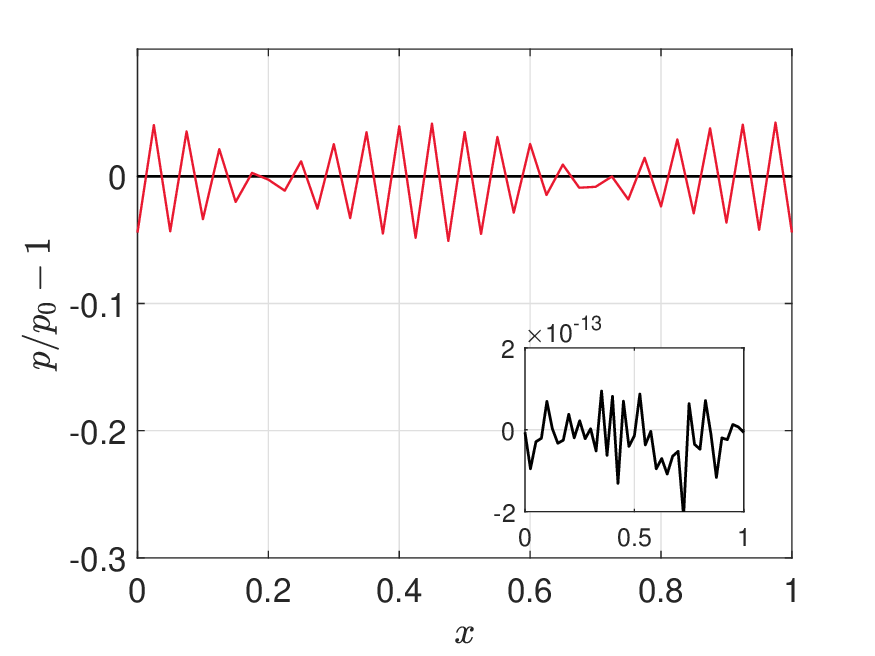}
    \caption{}
    \end{subfigure}
    \begin{subfigure}[b]{0.45\textwidth}
    \centering
    \includegraphics[width=\linewidth]{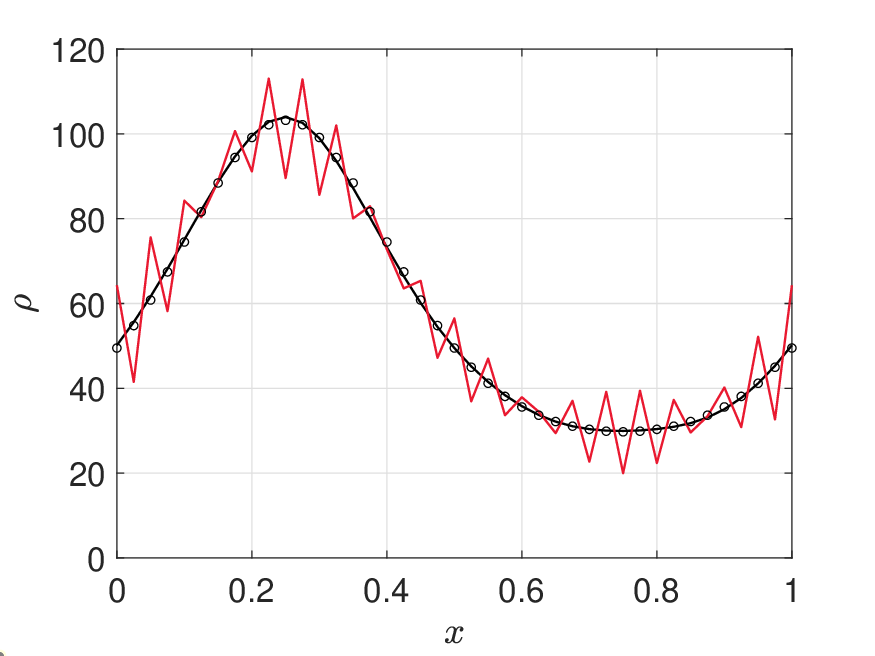}
    \caption{}
    \end{subfigure}
    \begin{subfigure}[b]{0.45\textwidth}
    \centering
    \includegraphics[width=\linewidth]{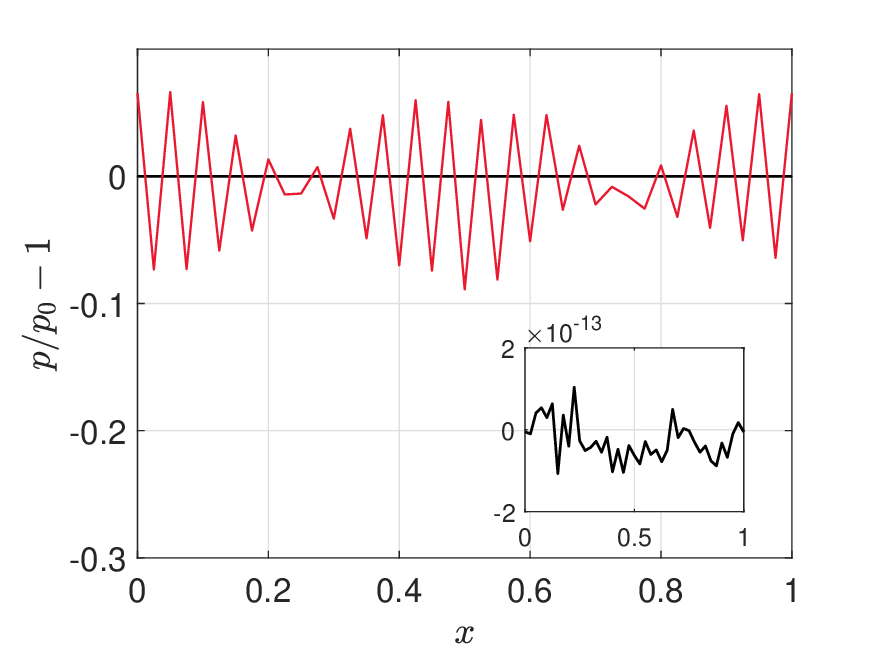}
    \caption{}
    \end{subfigure}
    \caption{Density and pressure profiles for the one-dimensional density wave test. (a)-(b): thermally perfect gas ($t=16$), (c)-(d) van der Waals equation of state in supercritical conditions ($t=13$), (e)-(f) Peng--Robinson equation of state in supercritical conditions ($t=14$).}
    \label{fig:DW_rho_p}
\end{figure}
The one-dimensional density wave is solved in the domain $\Omega:x\in[0,L]$ with $L=1$, discretized in $N=41$ evenly spaced points ($h=0.025$).
A fourth-order accurate spatial discretization is employed.
To minimize contamination from time-integration errors, the CFL number has been set to \num{5e-3} for all the considered cases.
The initial conditions correspond to a smooth density perturbation convected at constant velocity, defined as
\begin{equation}
    \begin{cases}
        \rho(x,0) & = \rho_0\left[A + Be^{\sin\left(\frac{2\pi x}{L}\right)}\right]\\
        u(x,0) & = u_0\\
        p(x,0) & = p_0
    \end{cases}
\end{equation}
with $\{\rho_0,p_0\}=\{\rho_c,100\}$ for the van der Waals and Peng--Robinson models in the supercritical regime, while $\{\rho_0,p_0\}=\{\rho_{\mathrm{SATP}},0.45\}$ for the thermally perfect one, where $\rho^*_{\mathrm{SATP}}=1.795$ Kg/m$^3$. Modulation constants are $A=0.07$ and $B=0.12$.
The value $u_0=1$ sets the reference time $t_{\mathrm{ref}}=L/u_0=1$, hence $t=t^*$.

To monitor the onset of numerical instabilities, we track the normalized variation of a generic quantity $\phi$ defined as
\begin{equation*}
\langle \phi \rangle = \frac{\int_{\Omega} \phi \, \dd\Omega - \int_{\Omega} \phi_0 \, \dd\Omega}{\int_{\Omega} \phi_0 \, \dd\Omega}.
\end{equation*}
In particular, we consider the kinetic energy $\rho\kappa$. Since both pressure and velocity are expected to remain constant for this problem, the kinetic energy should be preserved up to machine precision. Therefore, deviations in $\langle \rho\kappa \rangle$ provide a sensitive indicator of scheme robustness.

Fig.~\ref{fig:DW_rhok} reports the evolution of $\langle \rho\kappa \rangle$ for the schemes under consideration. All shown non-PEP schemes exhibit numerical blow-up. Among them, the first to become unstable is KEEP, which is not PEP even for ideal gases. This is followed by KEEP$_{\text{PE}}$, which satisfies the PEP property only for calorically perfect gases, and by APEC, which is only approximately PEP for real-gas models. In contrast, EPEP-RG maintains the error on kinetic energy at machine-zero throughout the simulation. The APEP-RG scheme (not displayed) has performances almost indistinguishable from that of the EPEP-RG in this time interval. These trends are consistent across all the tested thermodynamic models: thermally perfect, van der Waals, and Peng--Robinson.
\begin{figure}[tb]
    \centering    
    \subfloat[]{\includegraphics[width=0.5\linewidth]{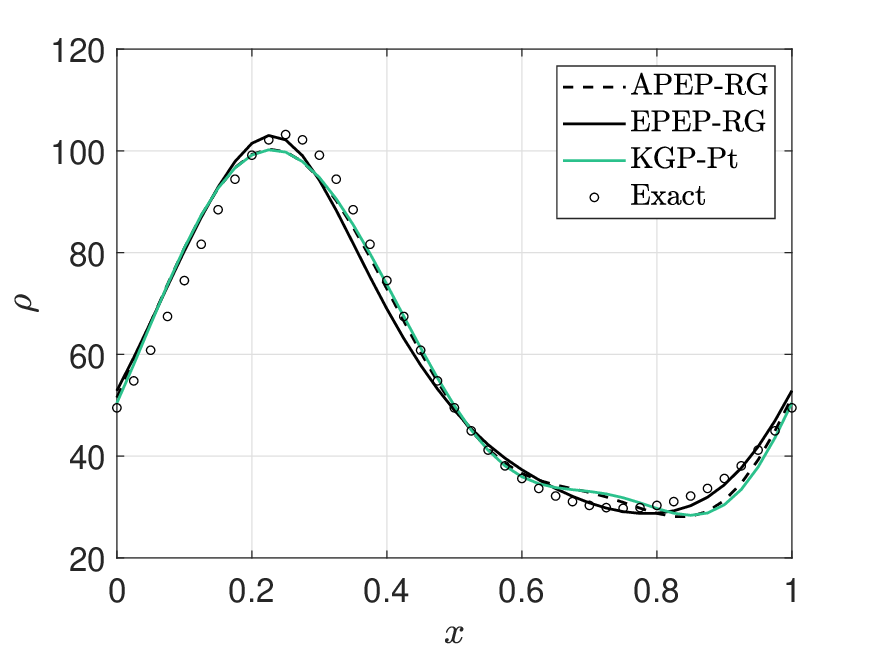}}    \subfloat[\label{fig:DW_t100_EPEPvsAPEP_rhoE}]{\includegraphics[width=0.5\linewidth]{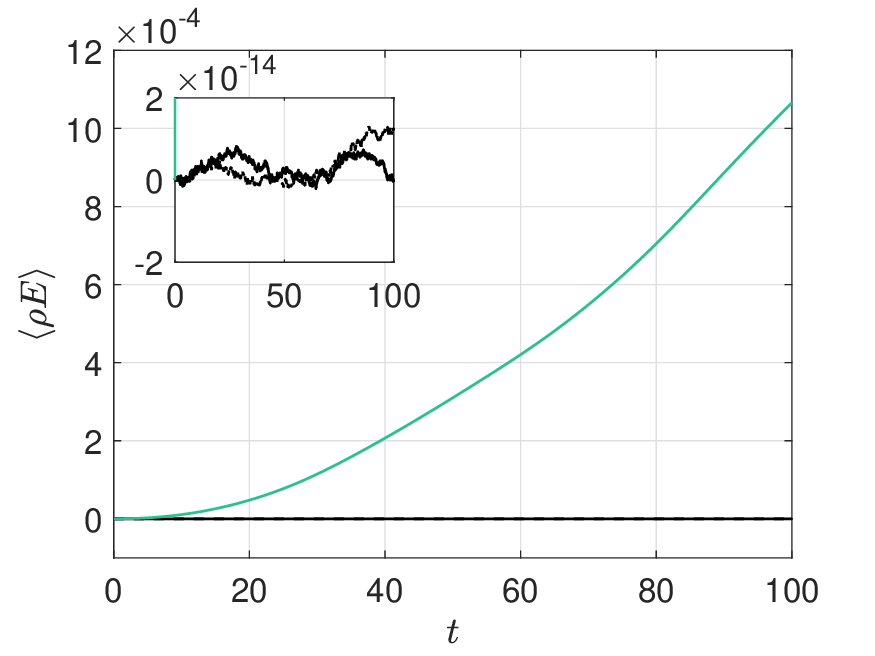}}
    \caption{Density profile (a) and total energy evolution (b) for the one-dimensional density wave test for the Peng--Robinson equation of state in supercritical conditions at $t=100$.}
    \label{fig:DW_PR_t100}
\end{figure}

Further insight is provided in Fig.~\ref{fig:DW_rho_p}, which shows solution snapshots at a time when KEEP and KEEP$_{\text{PE}}$ have already become unstable, while APEC is still running. For EPEP-RG, the pressure remains equal to its initial value and the density closely matches the exact solution. On the other hand, APEC exhibits spurious oscillations in both pressure and density, indicating the onset of numerical degradation despite not having blown up yet.

For the Peng--Robinson model, the EPEP-RG, APEP-RG and KGP$_{\text{Pt}}$ simulations have been performed up to a final time $t=100$, without recording any instability. In Fig.~\ref{fig:DW_PR_t100}, the density profile solution is reported at $t=100$, together with the time history of the total energy. Interestingly enough, the density profiles of the APEP-RG and KGP$_\text{Pt}$ formulations show similar dispersion features, a circumstance that should probably be attributed to the fact that the two formulations share the same mass and momentum fluxes. Fig.~\ref{fig:DW_t100_EPEPvsAPEP_rhoE} shows that, as predicted, the global total energy remains constant up to machine precision in this long simulation for the EPEP-RG and APEP-RG schemes, whereas for the KGP$_\text{Pt}$ scheme a slow accumulation is present during the whole simulation, reaching a final value of the order $10^{-3}$.

\begin{figure}[tb]
    \centering    
    \subfloat[]{\includegraphics[width=0.5\linewidth]{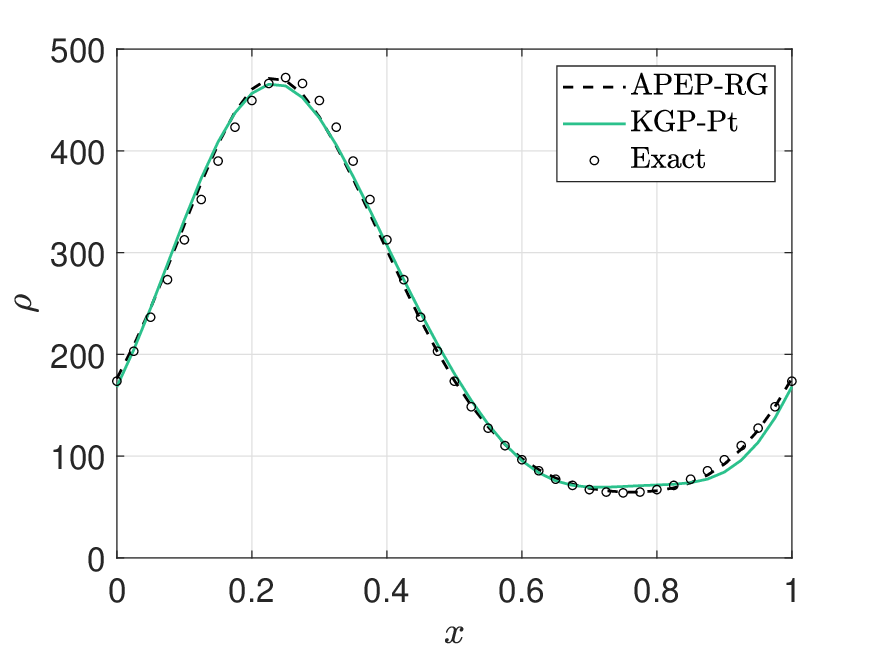}}
    \subfloat[]{\includegraphics[width=0.5\linewidth]{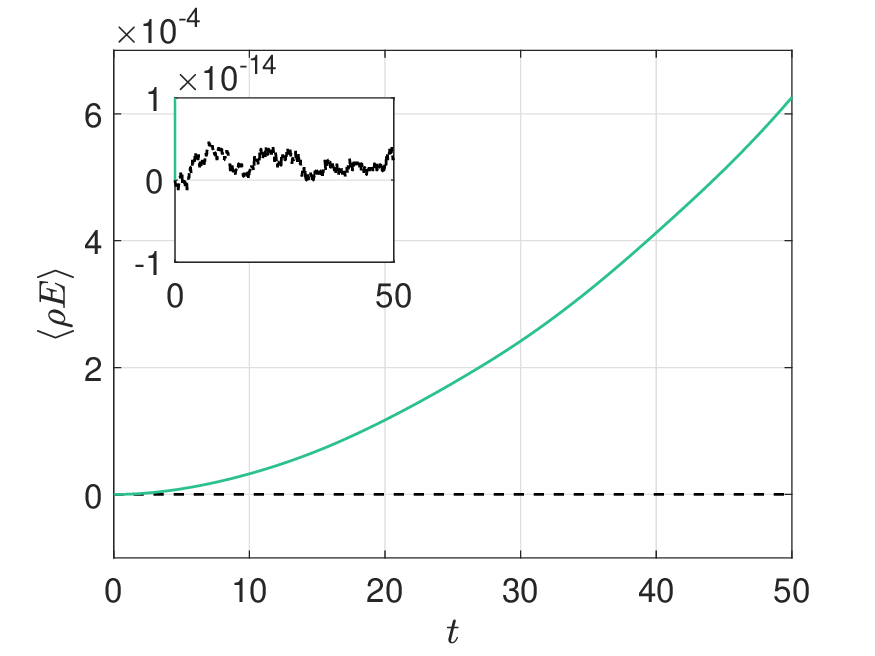}}
    \caption{Density profile (a) and total energy evolution (b) for the one-dimensional density wave test for the Peng--Robinson equation of state in transcritical conditions at $t=50$.}
    \label{fig:DW_PR_Transcritical_t50}
\end{figure}

To assess the performances of the newly proposed schemes in transcritical conditions, a simulation has been also carried out with the Peng--Robinson model for initial conditions $\{A,B,\rho_0,p_0\}=\{0,2/3,\rho_c,135\}$. In this transcritical case, in addition to the KEEP, APEC and KEEP$_{\text{PE}}$ formulations, also the EPEP-RG scheme shows strong instabilities, probably due to the near-singular behavior of the thermodynamic derivatives in the vicinity of the pseudo-critical line, where the main thermodynamic quantities---such as internal energy or the speed of sound---undergo rapid and large variations that can violate the assumptions on which the exact discrete formulation relies. 
Nevertheless, the APEP-RG formulation shows a good robustness,  similar to that of the  KGP$_{\text{Pt}}$ scheme, which is exactly PEP for this case.
Note that for this transcritical case, which is particularly susceptible to instabilities, even the KGP$_{\text{Pt}}$ scheme eventually diverges at long times. 
In fact, a long simulation shows blow up for both the APEP-RG and KGP$_{\text{Pt}}$ schemes at $t\approx 96$ and $t\approx 107$, respectively. Consequently, it seems that no numerical method currently available is capable of handling this test for sufficiently long times.
This result could possibly be consistent with the previous observations about the instability near the pseudo-critical line as a consequence of the direct computation of the thermodynamic derivatives to solve the main system of equations. In fact, the KGP$_{\mathrm{Pt}}$ involves the expression of the speed of sound, whose evaluation inherently requires several such derivatives, making the scheme susceptible to some ill-conditioning in the transcritical region.
In Fig.~\ref{fig:DW_PR_Transcritical_t50}, APEP-RG and KGP$_{\text{Pt}}$ density profiles and total energy evolutions are reported at $t=50$. APEP-RG  shows a slightly better agreement with the exact solution, while KGP$_{\text{Pt}}$ exhibits a lack of total energy conservation, with an error accumulating during the evolution and reaching a final value of approximately \num{6e-4}.

\subsection{Two-dimensional inviscid double-jet flow at high-enthalpy and supercritical conditions}\label{sec:test_dj}
The two-dimensional, inviscid double-jet flow is simulated in the rectangular domain $(x,y)\in[0,L]\times[-L/4,L/4]$, with $L=1$, discretized in $N_x\times N_y=65\times 33$ evenly spaced points with second-order accurate spatial discretizations, with CFL set as $0.01$ to constrain time-integration errors.
The flow is initialized as
\begin{equation}
    \begin{cases}
        \begin{cases}
            u(x,y,0) & = A_u\left[1 + A_u\tanh{\left(\theta(y+L/10)\right)}\right]\\
            T(x,y,0) & = aA_t\left[3/2 - A_t\tanh{\left(\theta(y+L/10)\right)}\right]
        \end{cases}
        \qquad \text{for}\quad y\leq 0\\
        \begin{cases}
            u(x,y,0) & = A_u\left[1 - A_u\tanh{\left(\theta(y-L/10)\right)}\right]\\
            T(x,y,0) & = aA_t\left[3/2 + A_t\tanh{\left(\theta(y-L/10)\right)}\right]
        \end{cases}
        \qquad \text{for}\quad y>0\\
        v(x,y,0)  = \varepsilon\sin{(2m\pi x/L)}\\
        p(x,y,0)  = p_0
    \end{cases}
\end{equation}
with the velocity field being the same for each gas model, given $\{A_u,\varepsilon,m\}=\{1/2,0.05,3\}$, to have a transversal jet that generates three roll-up vortices. Temperature and pressure fields are, on the other hand, established with respect to the specific equation of state. Thus, we set $\{a,A_t,p_0\}=\{2.6,2/3,0.1\}$ for the thermally perfect case and $\{a,A_t,p_0\}=\{2.5,1/2,150\}$ for the supercritical test, carried out by means of the van der Waals model. Finally, the parameter $\theta=30$ represents the thickness of the shear layer, and a reference time is defined as $t_{\mathrm{ref}}=m^{-1}/\max_{x,y}u(x,y,0)\approx 0.445$.

\begin{figure}[tb]
    \centering
    \subfloat[\label{fig:DoubleJet_vdW_x}]{\includegraphics[width=0.5\linewidth]{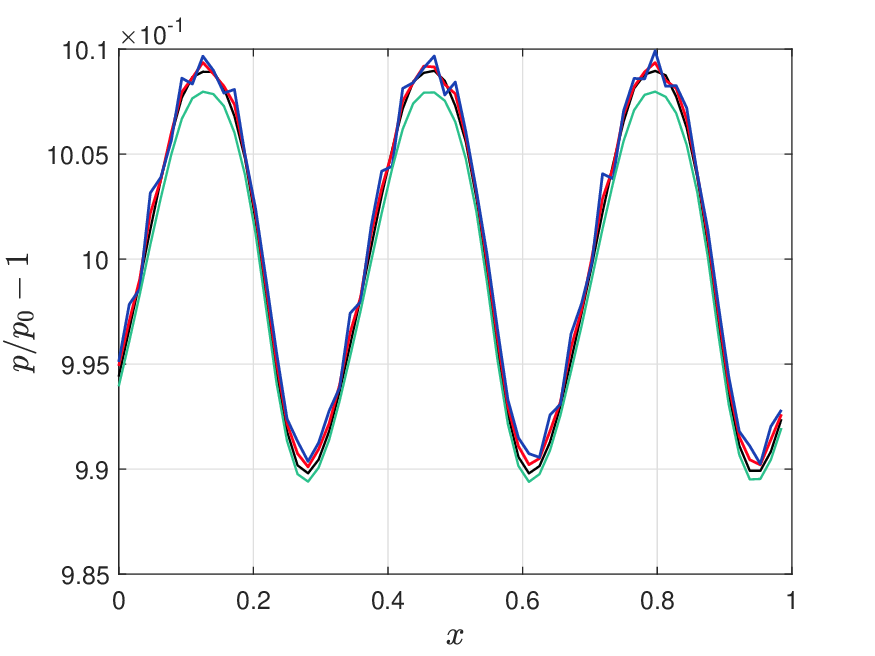}}
    \subfloat[\label{fig:DoubleJet_vdW_y}]{\includegraphics[width=0.5\linewidth]{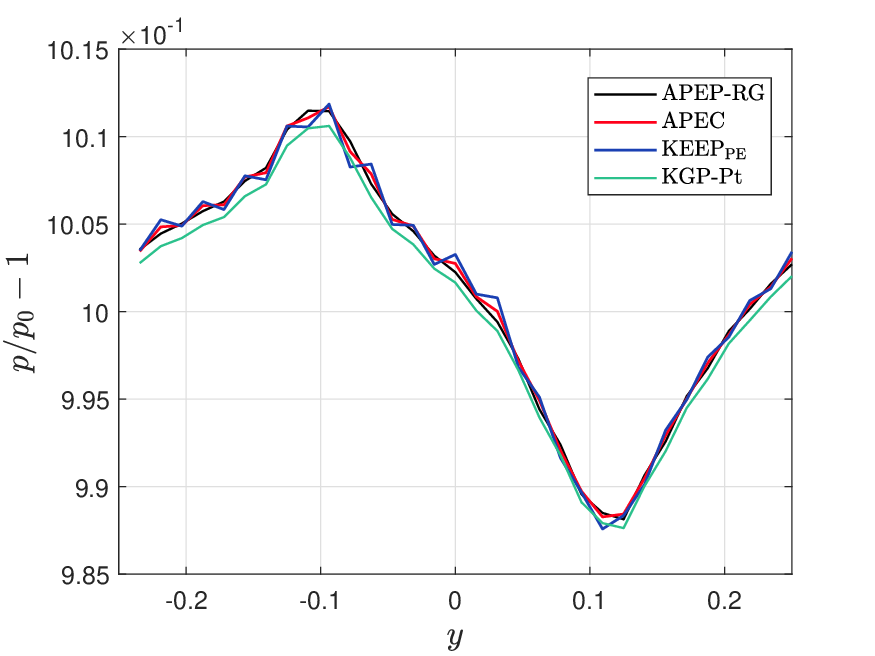}}\\
    \subfloat[\label{fig:DoubleJet_vdW_rhoEtot}]{\includegraphics[width=0.5\linewidth]{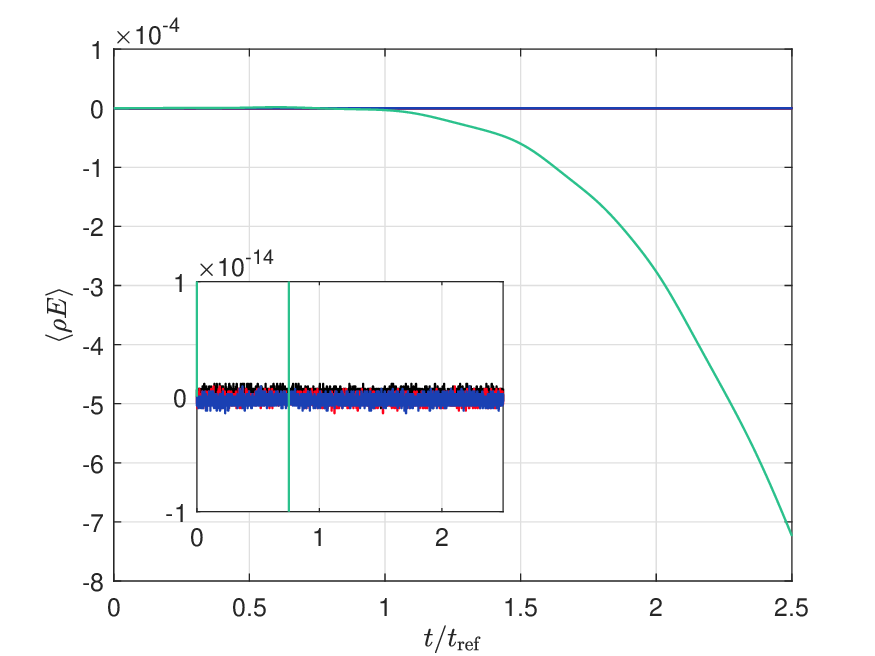}}
    \subfloat[\label{fig:DoubleJet_vdW_rhos}]{\includegraphics[width=0.5\linewidth]{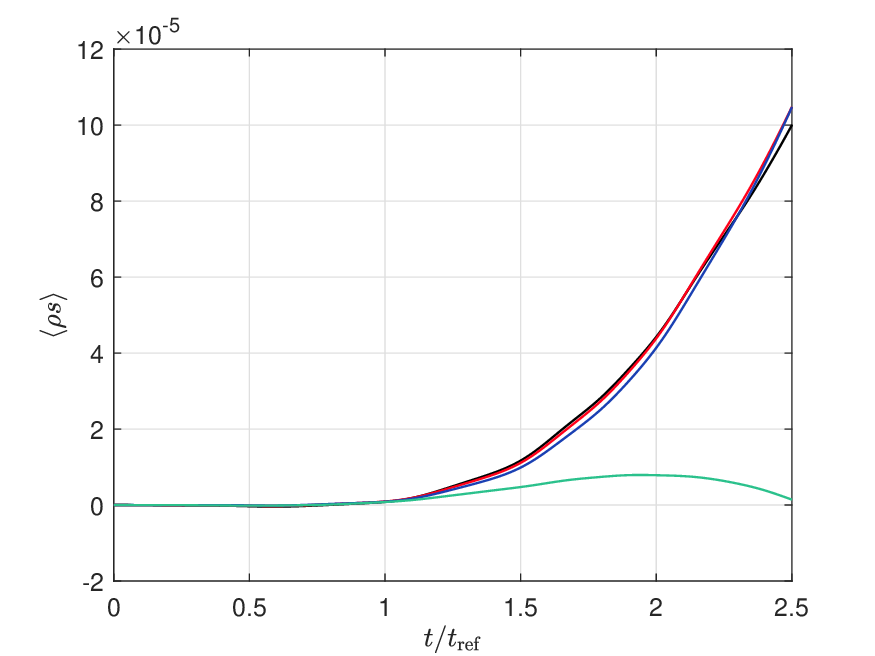}}\\
\caption{Two-dimensional inviscid double-jet flow for a van der Waals gas at supercritical conditions. (a) pressure profiles evaluated at $y=0.33$ and $t/t_\text{ref}=2.5$, (b) pressure profiles evaluated at $x=0$ and $t/t_\text{ref}=2.5$, (c) total energy evolution, (d) entropy evolution.}\label{fig:DoubleJet_vdW}
\end{figure}

As the numerical mass flux defined in Eq.~\eqref{eq:PEP_Fluxes_hat} is singular in the wide, initially uniform regions, we decided to use only the APEP-RG formulation for this test case, compared against the KEEP$_\text{PE}$, APEC, and KGP$_\text{Pt}$ schemes, which have shown better performances in the previous 1D test case with respect to the KEEP scheme.
\begin{figure}[tb]
    \centering
    \subfloat[\label{fig:DoubleJet_TP_x}]{\includegraphics[width=0.5\linewidth]{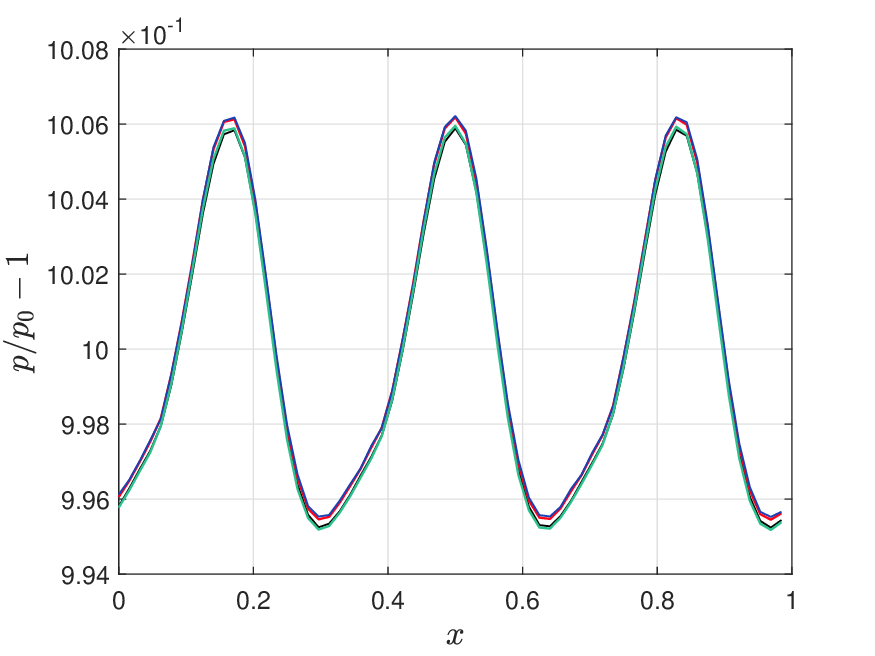}}
    \subfloat[\label{fig:DoubleJet_TP_y}]{\includegraphics[width=0.5\linewidth]{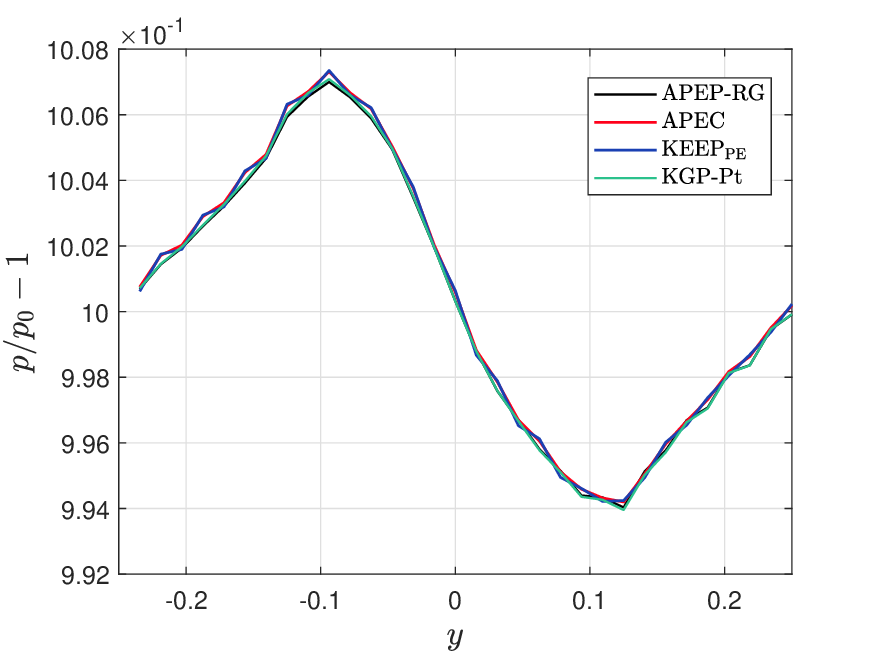}}\\
    \subfloat[\label{fig:DoubleJet_TP_rhoEtot}]{\includegraphics[width=0.5\linewidth]{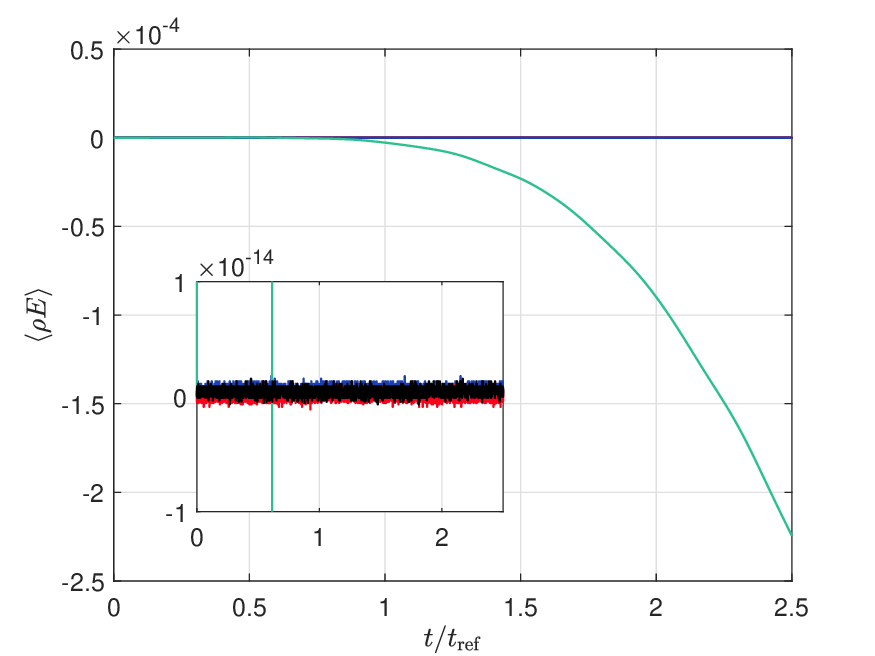}}
    \subfloat[\label{fig:DoubleJet_TP_rhos}]{\includegraphics[width=0.5\linewidth]{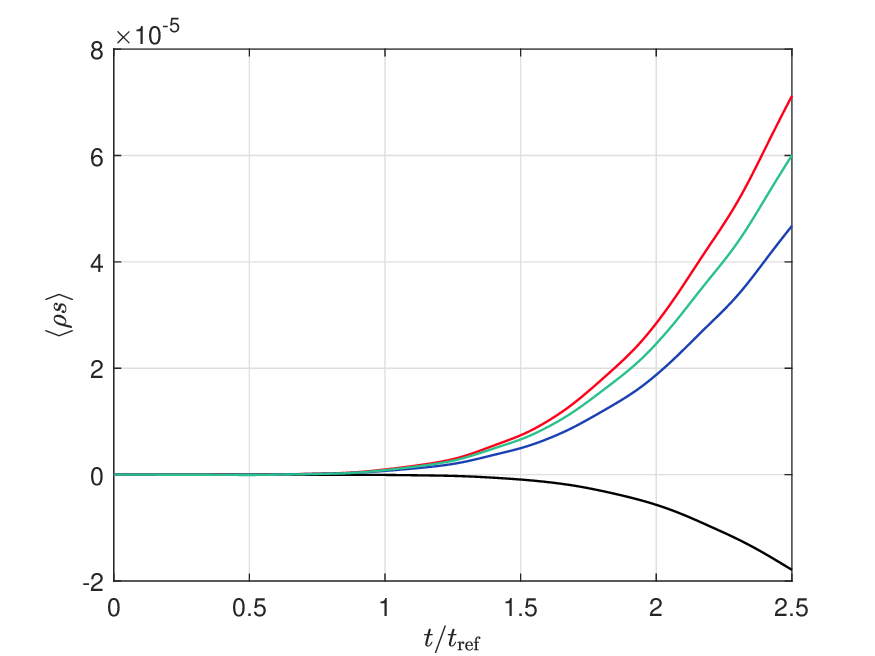}}\\

\caption{Two-dimensional inviscid double-jet flow for a thermally perfect gas. (a) pressure profiles evaluated at $y = 0.33$ and $t/t_\text{ref}=2.5$, (b) pressure profiles evaluated at $x=0$ and $t/t_\text{ref}=2.5$, (c) total energy evolution, (d) entropy evolution.}\label{fig:DoubleJet_TP}
\end{figure}
In Fig.~\ref{fig:DoubleJet_TP}, the results of the high-enthalpy case are reported, in terms of pressure profiles along horizontal and vertical lines at $y=0.33$ and $x = 0$, respectively (Fig.~\ref{fig:DoubleJet_TP_x} and \ref{fig:DoubleJet_TP_y}), together with the time evolution of the total energy and entropy. Although the results are generally satisfactory for all the schemes considered, Fig.~\ref{fig:DoubleJet_TP_y} shows that KEEP$_\text{PE}$ and APEC are starting to exhibit point-to-point oscillations in the vertical profile of pressure. The total energy evolution depicted in Fig.~\ref{fig:DoubleJet_TP_rhoEtot}confirms that the KGP$_\text{Pt}$ scheme steadily deviates from exact conservation, whereas the global entropy evolution reported in Fig.~\ref{fig:DoubleJet_TP_rhos} shows that all the considered formulations violate the exact entropy preservation, as none of them is exactly entropy conservative.
The pressure oscillations highlighted for the KEEP$_\text{PE}$ and APEC schemes in Fig.~\ref{fig:DoubleJet_TP_y} are much more visible in the supercritical case computed with the van der Waals model, reported in Fig.~\ref{fig:DoubleJet_vdW_x} and \ref{fig:DoubleJet_vdW_y}. In this simulation the APEP and KGP$_\text{Pt}$ schemes remain essentially free of oscillation.
\subsection{Two-dimensional inviscid double-jet flow at transcritical conditions}\label{sec:test_dj_trans}

\begin{figure}[tb]
    \centering
    \subfloat[\label{fig:DoubleJet_PR_x}]{\includegraphics[width=0.5\linewidth]{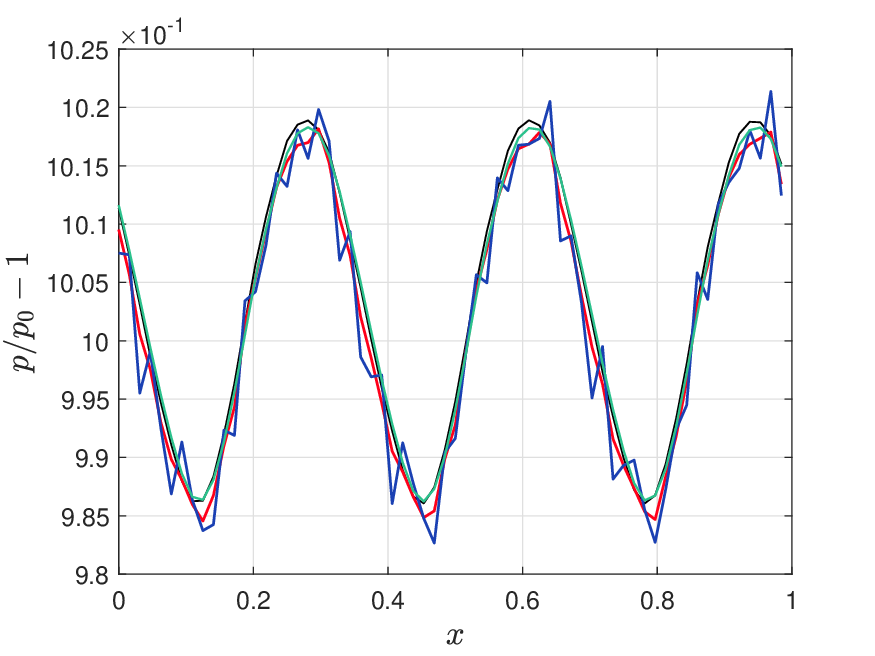}}
    \subfloat[\label{fig:DoubleJet_PR_y}]{\includegraphics[width=0.5\linewidth]{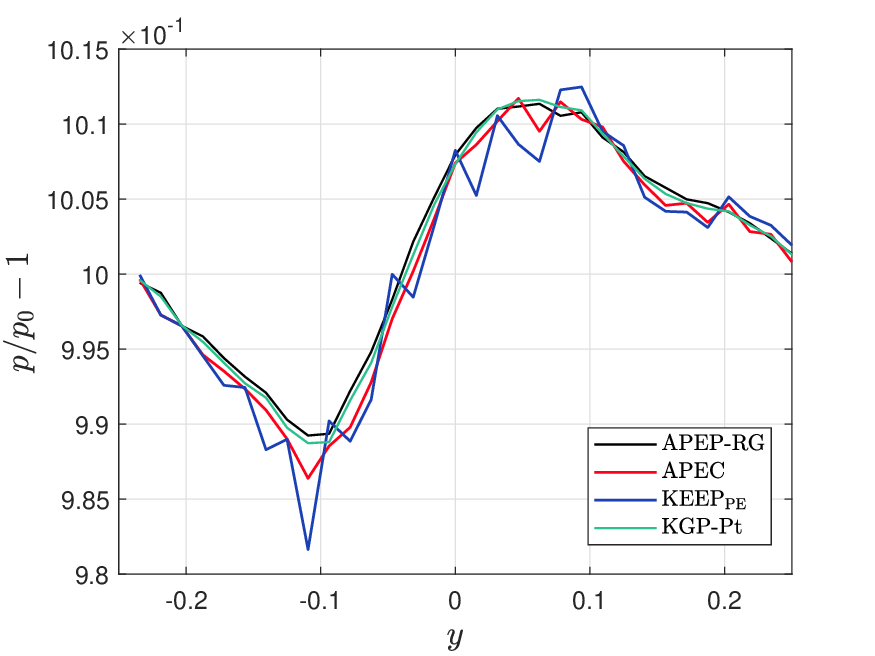}}\\
    \subfloat[\label{fig:DoubleJet_PR_rhoEtot}]{\includegraphics[width=0.5\linewidth]{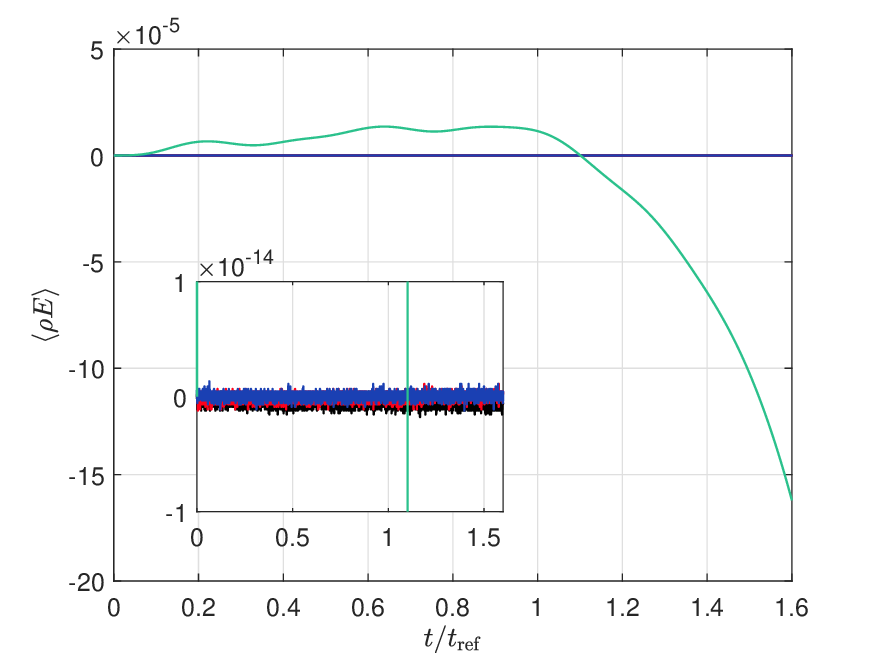}}
    \subfloat[\label{fig:DoubleJet_PR_rhos}]{\includegraphics[width=0.5\linewidth]{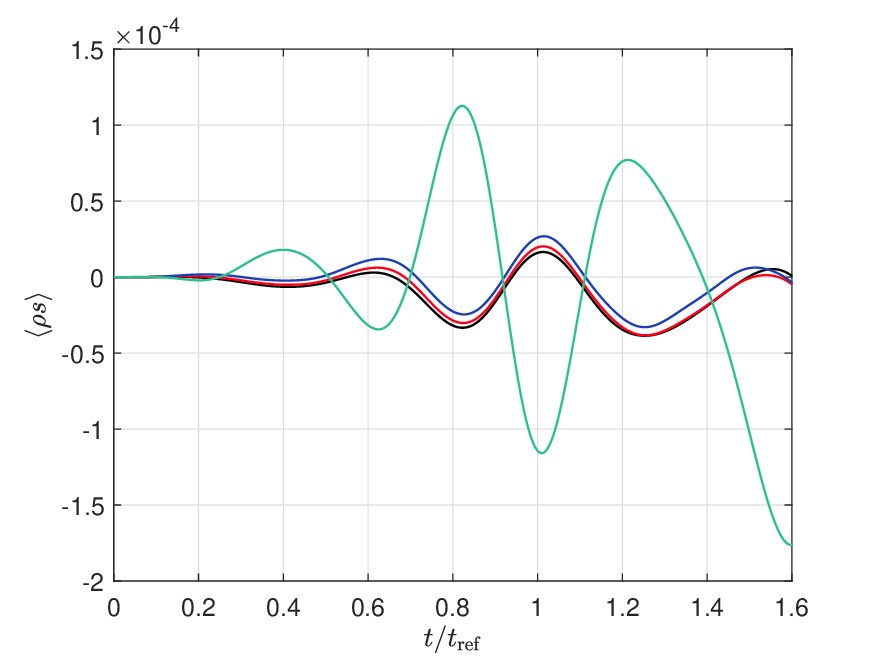}}
    
    \caption{Two-dimensional inviscid double-jet flow for a Peng--Robinson gas at transcritical conditions. (a) pressure profiles evaluated at $y = 0.33$ and $t/t_\text{ref}=2.5$, (b) pressure profiles evaluated at $x=0.5$ and $t/t_\text{ref}=2.5$, (c) total energy evolution, (d) entropy evolution. }\label{fig:DoubleJet_PR}
\end{figure}

\begin{figure}[tb]
    \centering
    \subfloat[APEP-RG]{\includegraphics[width=0.5\linewidth]{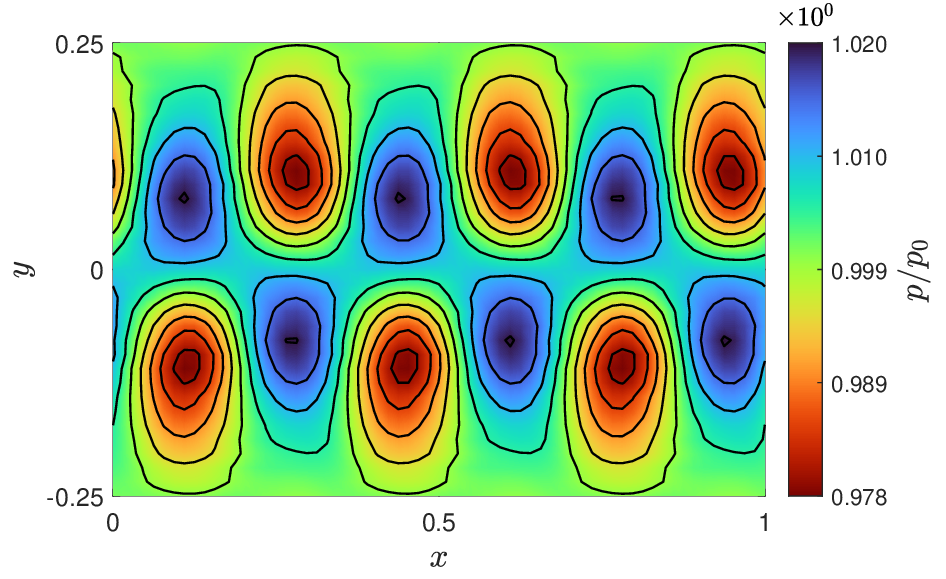}}
    \subfloat[KGP$_\text{Pt}$]{\includegraphics[width=0.5\linewidth]{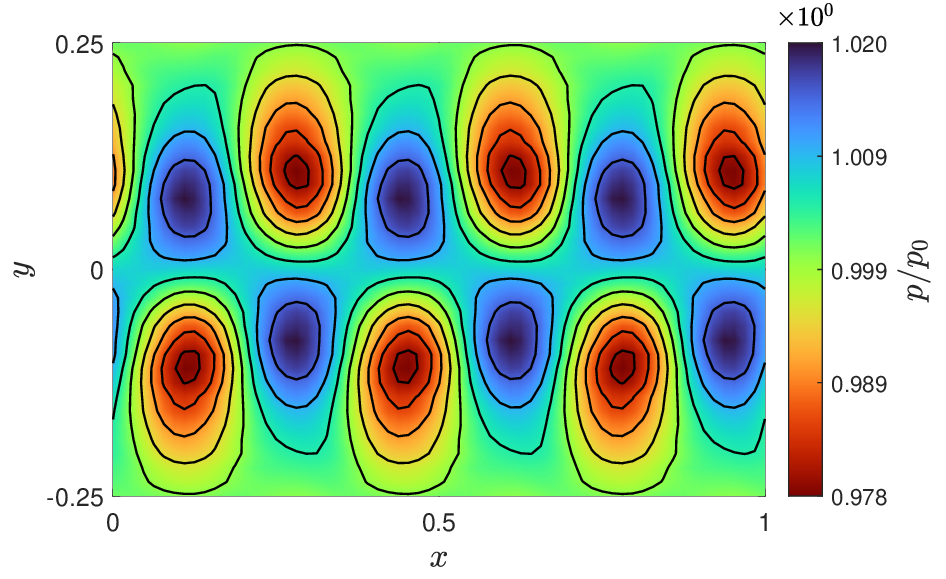}}
    \\
    \subfloat[APEC]{\includegraphics[width=0.5\linewidth]{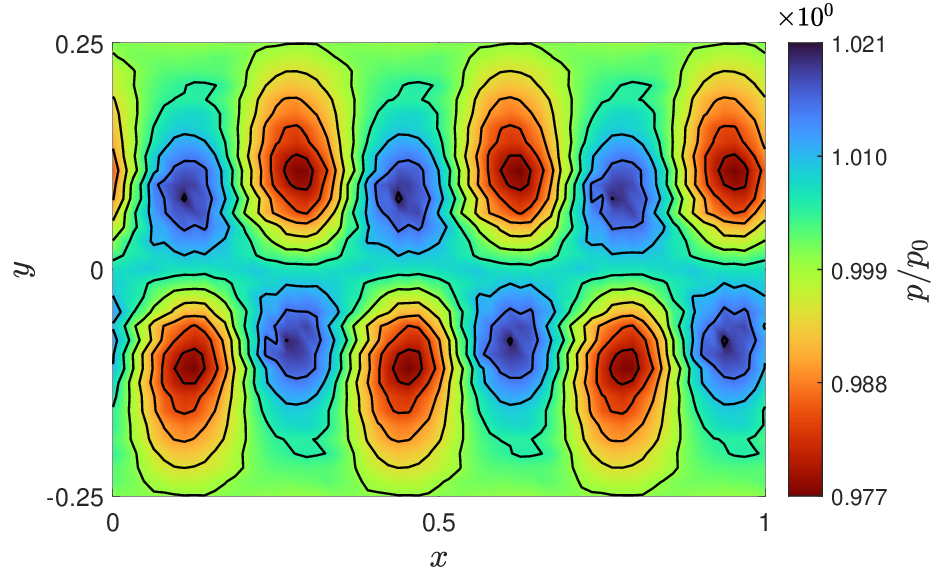}}
    \subfloat[KEEP$_{\mathrm{PE}}$]{\includegraphics[width=0.5\linewidth]{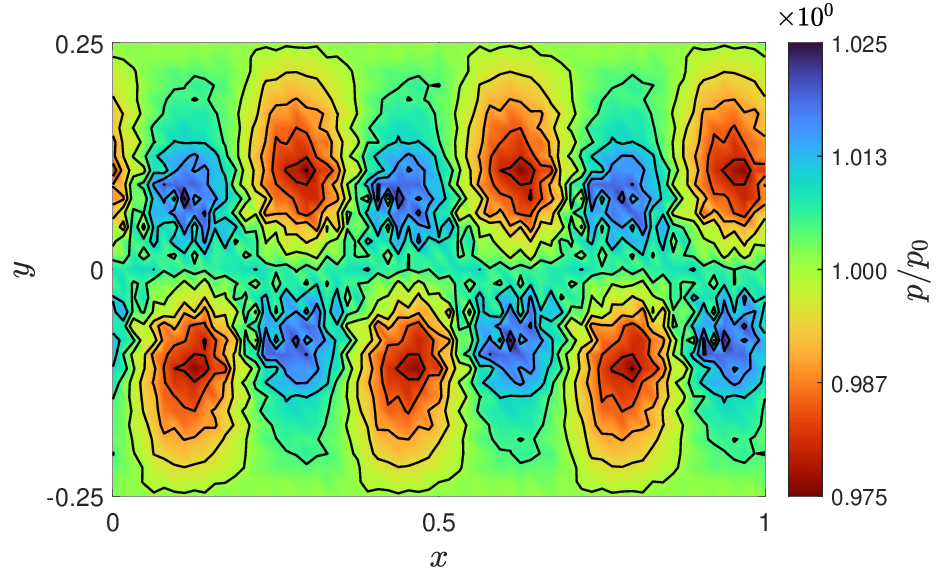}}
    \\
    \caption{Instantaneous pressure fields for the two-dimensional inviscid double-jet flow for a Peng--Robinson gas at transcritical conditions and $t/t_{\mathrm{ref}}=1.6$.}\label{fig:DoubleJet_PR_PressureFields}
    \end{figure}

Transcritical conditions necessitate a more detailed analysis due to the different behavior of thermodynamic derivatives near the critical point, especially when discretizing the energy equation which requires the computation of internal energy and its gradients in the thermodynamic space. In this section, the inviscid double-jet flow is simulated onto the same grid and within the same numerical setup presented in Section~\ref{sec:test_dj}. Initial conditions also share the same symbolic form, with the $\{a,A_t,p_0\}=\{2,1/2,180\}$. This ensures a dimensional temperature $T^*\in[{\sim}\, 298.8,{\sim}\,587.9] \,\si{\kelvin}$, therefore crossing the critical temperature for CO$_2$, $T^*_c=\SI{304.12}{\kelvin}$. Corresponding dimensional pressure is $p^*\approx 2.43\times p^*_c=2.43\times \SI{73.8}{\atmosphere}$. This time, the Peng--Robinson model has been used to carry out the simulations.
Figures~\ref{fig:DoubleJet_PR_x} and \ref{fig:DoubleJet_PR_y}  report the usual pressure profiles at $t = 1.6$. In this case, the pressure oscillations are much more evident for the KEEP$_\text{PE}$ and APEC schemes, even if the simulation is stopped at an earlier time. The total energy evolution is consistent with the other simulations, whereas the entropy evolution reported in Fig.~\ref{fig:DoubleJet_PR_rhos} shows an oscillating behavior, which is exacerbated  for the KGP$_\text{Pt}$ formulation. 
Fig.~\ref{fig:DoubleJet_PR_PressureFields} reports a snapshot of the two-dimensional pressure field as calculated by the various formulations, confirming the contamination of the solution due to the growing oscillations in the KEEP$_\text{PE}$ and APEC schemes.

\section{Conclusions}\label{sec:conclusions}
As highlighted throughout this work, the numerical simulation of compressible flows for non-ideal fluids is frequently challenged by the generation of spurious pressure oscillations. While addressing this issue requires numerical methods that satisfy the pressure-equilibrium-preserving (PEP) property, ensuring strict PEP compliance for general equations of state has historically proven difficult. Previous attempts often compromise the exact discrete conservation of total energy, a property that remains essential for maintaining physical consistency and correctly capturing shocks.

In this work, we demonstrate that the exact discrete conservation of mass, momentum, and total energy is not mutually exclusive with the PEP condition. Although the present mathematical framework is derived in the context of finite-difference discretizations, the resulting two-point numerical fluxes are highly versatile. They can be seamlessly applied to other spatial discretization frameworks, such as structured and unstructured finite volume or discontinuous Galerkin formulations.

Our main contribution is the development of a fully conservative and exactly pressure-equilibrium-preserving scheme, denoted as Exact PEP (EPEP-RG). We derive a generic formula applicable to any arbitrary equation of state and provide specific, computationally viable flux formulations for thermally perfect, van der Waals, and Peng--Robinson gases.
Building upon this theoretical foundation, and recognizing the complexities inherent to certain thermodynamic regimes, we also propose a robust practical alternative: the Approximate PEP (APEP-RG) scheme. The APEP-RG formulation strictly maintains full primary conservation, including kinetic-energy preservation by convection, while enforcing the PEP condition in an approximate sense.

We have validated the proposed schemes using rigorous numerical benchmarks. In the density wave advection test, the EPEP-RG scheme is able to successfully preserve pressure equilibrium across a variety of EoS. This behavior contrasts favorably with standard formulations from the literature, which either become unstable or fail to conserve total energy. Furthermore, simulations of a compressible mixing layer confirm the exact conservation of total energy alongside the elimination of spurious oscillations in the pressure field, effectively addressing the energy conservation limitations of previous non-oscillatory schemes.
While the exact EPEP-RG formulation can encounter numerical singularities due to problematic thermodynamic derivatives in transcritical regimes, the APEP-RG scheme reliably circumvents these issues. Overall, the proposed schemes demonstrate highly favorable robustness, stability, and accuracy. The APEP-RG framework, in particular, emerges as a simple and resilient tool capable of managing severe thermodynamic nonlinearities---including transcritical phenomena---without sacrificing primary conservation or numerical stability.

Despite these successes, certain limitations remain that pave the way for future research. The EPEP-RG scheme currently requires a specialized fix to properly handle the singularities of thermodynamic derivatives across phase boundaries in transcritical regimes or in areas where thermodynamic quantities are nearly uniform. The APEP-RG scheme provides a robust workaround for these issues, while also introducing a degree of flexibility regarding the choice of the density averaging operator. Future work could systematically explore the characteristics of various averaging operators to identify the optimal formulation, potentially exploiting this degree of freedom to impart additional structure-preserving properties to the scheme, such as discrete entropy conservation or stability.

Further efforts will focus on extending this framework to multi-component and multi-phase flow formulations, as well as evaluating the schemes' performances on more complex, physically realistic configurations, such as wall-bounded turbulent flows. Ultimately, the conservative and equilibrium-preserving methodologies introduced herein represent a significant step forward toward a unified, robust framework for the high-fidelity simulation of complex real-gas flows.

\bibliographystyle{model1-num-names}
\bibliography{Biblio_KEP_Compr}

\end{document}